\documentclass[epj]{svjour}
\usepackage{amsmath}
\usepackage{amsopn}
\usepackage{xcolor}
\usepackage{graphicx,subfigure}
\usepackage{color,soul}
\usepackage{mathtools, cuted}
\usepackage{graphicx}
\usepackage{subfigure}
\usepackage[breaklinks=true,colorlinks=true,linkcolor=blue,urlcolor=blue,citecolor=blue]{hyperref}
\usepackage{cite}

	\newcommand{\ket}[1]{\left| #1 \right\rangle}
	\newcommand{\bra}[1]{\left\langle #1 \right|}

	\newcommand{\Tr}{\operatorname{Tr}}
\begin{document}
\title{Dissipative entanglement swapping in the presence of detuning and Kerr medium: Bell state measurement method}

\author{M Ghasemi\inst{1} \thanks{m.ghasemi@stu.yazd.ac.ir}
\and M K Tavassoly\inst{1,}\inst{2} \thanks{mktavassoly@yazd.ac.ir}
\and A Nourmandipour\inst{1} \thanks{anourmandip@gmail.com}
}                    

\institute{Atomic and Molecular Group, Faculty of Physics, Yazd University, Yazd  89195-741,
Iran \and The Laboratory of Quantum Information Processing, Yazd University, Yazd, Iran}
\date{Received: date / Revised version: date}
%
\abstract{
 In this paper, we investigate the possibility of the entanglement swapping between two independent nonperfect cavities consisting of an atom with finite lifetime of atomic levels (as two independent sources of dissipation) which interacts with a quantized electromagnetic field in the presence of detuning and Kerr medium. In fact, there is no direct interaction between the two atoms, therefore, no entanglement exists between them. We use the Bell state measurement performing on the photons leaving the cavities to swap the entanglement stored between the atom-fields in each cavity into atom-atom. Our motivation is arisen from the fact that two-qubit entangled states are of enough interest for quantum information science and technologies.  We discuss the effect of initial state of the system, detuning parameter, the Kerr medium and the two dissipation sources on the swapped entanglement to atom-atom.  We interestingly find that when the atomic decay rates and photonic leakages from the cavities are equal, our system behaves as an ideal system with no dissipation. Our results show that it is possible to create a long-living atom-atom maximally entangled state in the presence of Kerr effect and dissipation; we determine these conditions in detail and also establish the final atom-atom Bell state.
} 
\titlerunning{Dissipative entanglement swapping in the presence \dots}
\authorrunning{M Ghasemi, M K Tavassoly, A Nourmandipour}
\maketitle
\section{Introduction}\label{Introduction}

 Quantum entanglement as a nonlocal correlation between quantum systems is one of the most extraordinary aspects of the quantum theory with no classical elucidation \cite{benatti2003}. Due to its various applications such as quantum key distribution \cite{Lo2012}, quantum cryptography \cite{Gisin2002}, quantum teleportation \cite{Braunstein1995,Caruso2010,sehati2017}, superdense coding \cite{Mattle1996,Harrow2004}, sensitive measurements \cite{Richter2007}, and quantum telecloning \cite{Muarao1999},  this notion has been placed at the heart of quantum information processing. Recently, three-party remote state preparation schemes based on entanglement property has been considered in \cite{Zhou2014}.
   As a few proposals for the implementations in producing entangled states one may refer to  trapped ions \cite{Turchette1998} and two-coupled flux qubits \cite{Izmalkov2004}. It is well-known that the interaction of atoms with cavity field could also be considered as an efficient source of different types of entanglement, such as atom-atom, atom-field and field-field entanglement \cite{Liao2011,Faghihi2012,Solano2002}. Such an interaction model which is called the Jaynes-Cummings model (JCM) \cite{Jaynes1963}, relies on the mutual coupling between a two-level atom and a single-mode field in the rotating wave approximation. The great advantage of this model has been concealed in its various possible generalizations. For instance, one could think of considering an arbitrary number of qubits  \cite{Faghihi2013}, multi-mode cavity field \cite{Faghihi2014}, multi-level atoms \cite{Baghshahi2014}, intensity-dependent coupling \cite{Faghihi2012}, moving atoms \cite{Faghihi2013}, etc.\\
   From the point of view of entanglement property, it is not necessary that subsystems directly interact with each other to be entangled, i.e., indeed systems distributed over long distances with no common past may become entangled. In such cases, one could think of entangling the subsystems constructing a more general system, with the help of two (or more) another entangled quantum subsystems.
     This is at the heart of entanglement swapping process \cite{Zukowski1993}, a fascinating phenomenon which plays an important role in stimulating curiosity to understand the quantum concepts \cite{Megidish2013}.
    This scheme has been originally proposed for entanglement swapping between a pairs of particles \cite{Zukowski1993}.
      Soon after, this notion was generalized to multi-particle quantum systems \cite{Bose1998} and also to the continuous variable systems \cite{Polkinghorne1999}. In \cite{Jia2004} the authors have investigated the experimental demonstration of  unconditional entanglement swapping for continuous variables. The possibility of optimization of entanglement purification via entanglement swapping has been studied in \cite{Shi2000}. Loss-resistant state teleportation and entanglement swapping using a quantum-dot spin in an optical microcavity has been studied in \cite{Hu2011}. Entanglement swapping in two independent JCM has been discussed in \cite{Liao2011}. By replacing the unknown state with an entangled state, entanglement swapping can be considered as a special example of quantum teleportation \cite{Lee2011}. Recently, quantum dual signature scheme based on coherent states with entanglement swapping has been investigated in \cite{Liu2016}.
      The basic concept concealed behind the quantum swapping is the Bell state measurement (BSM) approach \cite{Pakniat2017,Pakniat2016}. This notion can be thought as a projection operator which projects the state of fields into a Bell state and leaves qubits in an entangled state \cite{Deng2006}. We should emphasize that Bell state can be generally generated through various schemes (for instance see \cite{Orieux2013,Brida2007,Mishima2008}). \\
    On the other hand,  the advantages of Kerr medium on various properties of atom-field interacting systems have been previously reported. A Kerr medium can be generated via the propagation of a conveniently shaped Gaussian beam in a GRIN media by introducing a second-order correction to the paraxial propagation of the beam \cite{Soto2014}. Kerr medium is utilized to detect the nonclassical states of light \cite{Hillery1991}. To emphasize on its usefulness in the quantum information processing aspects, one may refer to the possibility of generation of entangled macroscopic quantum states via Kerr medium \cite{gerry1999}, entanglement concentration protocols for partially entangled photons \cite{Sheng2012}, quantum teleportation \cite{Vitali2000}, and quantum computation \cite{Pachos2000}. In addition, it has been put forward to explore the role of  Kerr medium on the generation of nonclassical features (specially entanglement) of quantum systems \cite{Vitali2000}. The influence of a Kerr-like medium on the evolution of field entropy and entanglement in a three-level atom 
    (two two-level atoms and a two-mode field) in an optical cavity  
    has been investigated in \cite{Abdel2000} (\cite{1555-6611-24-12-125203}). Authors in \cite{Sheng2008} have considered efficient polarization entanglement  with cross-Kerr nonlinearity and a way to distinguish the hyperentangled Bell
    states completely with the help of cross-Kerr nonlinearity has been studied in \cite{Sheng2010}. A standard Kerr medium generates an appropriate intensity-dependent phase shift through a coupling quadratic in the photon number \cite{Yurke1986}. Recently the notion of deformed Kerr and generalized deformed Kerr media as well as their physical properties have been outlined by one of us \cite{Honarasa2012}.\\
   In this paper we intend to study the possibility of entanglement swapping between two independent atom-field subsystems, however, now in the presence of the Kerr medium and either atomic or photonic dissipations.  In this regard, due to the realization facts, i.e., the finite Q-cavity and the finite lifetime of atomic levels dissipative atom-field interaction should be considered. Such a dissipative atom-field dynamics can also be described by JCM \cite{Vogel}. We consider each subsystem as a generalized JCM with a Kerr term, in which each dissipative atom off-resonantly interacts with its nonperfect cavity field surrounded by a Kerr medium. Therefore, by considering suitable initial states, we are able to obtain the exact time-evolved state vector of each  atom-field subsystem, independently. With the help of BSM performing on the fields leaving the cavities, we show that how the produced atom-field entanglement can be swapped to field-field and atom-atom, resulting, in particular, a final atom-atom entangled state. A few words seems to be necessary in relation to the Bell state measurement technique.
   This process, which is indeed a joint quantum-mechanical measurement of two two-level systems (either field or qubit), plays a significant role in quantum information science, since it determines that the two fields or qubits live in which of the four associated Bell states.
   If two two-level systems were not previously entangled, by projecting into a (maximally entangled) Bell state (according to the projection rule of quantum measurements) they become entangled; therefore, a Bell measurement is really an entangling operation. Accordingly, the Bell state measurement is the inseparable step of quantum teleportation as well as the entanglement swapping protocols (an experimental setup realization of the model is depicted in Fig. \ref{fig:model}). We then quantify the amount of entanglement which is swapped via concurrence \cite{Wootters1998}. Finally, the influence of initial state of the atom-field, detuning parameter and the Kerr medium on the atom-atom entanglement in the presence and absence of dissipation sources are examined.  The noticeable advantage of this work lies in the preparation of long-living maximally atom-atom entangled state in the presence of Kerr medium and dissipation using the entanglement swapping for most of the times. Our results show that, the detuning parameter in the absence of losses decreases the maximum value of swapped entanglement in each period of time. However, in the presence of losses the atom-atom state converts to the maximally entangled state. Also, our system interestingly converts to the ideal system when the atomic decay rates from excited state of atoms and photon escape rates from the cavities are being equal.

   The rest of paper is organized as follow: In Sec. 2 we describe the system under consideration and obtain the state vector of the entire system at any time $t$. Then performing the BSM, we obtain the atom-atom quantum state.  In Sec.  3 we investigate the role of various parameters on the swapped entanglement to the two atoms. Finally, the paper ends with a summary and conclusion in Sec. 4.

\section{Model and its solution}\label{sec.model}

 \begin{figure}[ht]
    \centering
 \includegraphics[width=0.75\textwidth]{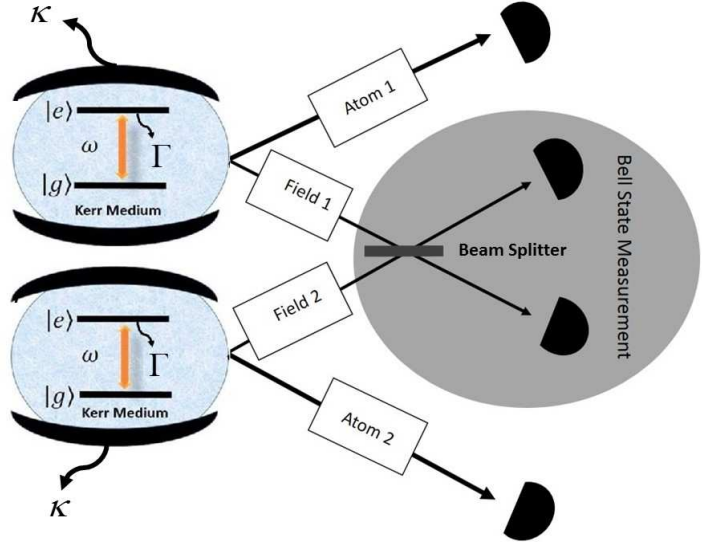}
    \caption{\label{fig:model}  The experimental realization of the entanglement swapping model. Each dissipative atom has been placed in its own nonperfect cavity in the presence of the Kerr medium. The Bell state measurement process which is performed on the photons leaving the cavities establishes entanglement generation between the two atoms. Therefore, entanglement is swapped to atom-atom from two noninteracting atom-field states.}
   \end{figure}
 Our model consists of two similar but separate nonperfect cavities, each contains a two-level atom with excited (ground) state $\ket{e}$ ($\ket{g}$) interacting with a single-mode cavity field in the presence of a Kerr medium. In addition, the leakage of photons from the cavity and the finite lifetime of atomic levels as two different dissipation sources are taken into account \cite{Vogel}. Therefore, the Hamiltonian describing the $i$th atom-cavity subsystem in the rotating wave approximation can be written as $(\hbar=1)$:
 \begin{equation}\label{eq1}
 \hat{H}_{\left( \mathrm{AF}\right)_i }=\frac{\omega_i}{2}\hat{\sigma}_{z_i}+\nu_i\hat{a_i}^{\dagger}\hat{a_i}+g_i\left( \hat{a_i}\hat{\sigma_i}^++\hat{a_i}^{\dagger}\hat{\sigma_i}^-\right)+\chi_i\hat{a_i}^{\dagger 2}\hat{a_i}^2-i\frac{\kappa_i}{2}\hat{a_i}^{\dagger }\hat{a_i}-i\frac{\varGamma_i}{2} \hat{\sigma_i}^+\hat{\sigma_i}^- ,
 \end{equation}
 where $i=1,2$ and $\hat{a}_i$ ($\hat{a}_i^{\dagger}$) is the annihilation (creation) operator. Also, $\omega_i$ and $\nu_i$ are the frequencies of the two-level atomic transition and the $i$th cavity field, respectively. $\hat{\sigma_i}^+$ ($\hat{\sigma_i}^-$) and $\hat{\sigma}_{z_i}$ are the raising (lowering) and population inversion operators of the $i$th atom, respectively. In the above relation, $g_i$ implies the coupling coefficient between the qubit and the field, and $\chi_i$  is the third-order component of the electric susceptibility of the Kerr medium. Finally, the two last terms denote the loss rate (photon absorption and scattering) by the mirrors ($\kappa_i$) and the atomic decay rate ($\varGamma_i$) from the exited states, respectively.  We suppose that the two atom-field subsystems are initially in the following separable state:
 \begin{eqnarray}\label{initialstate}
    \ket{\Psi(0)}&=&\ket{\Psi_{\mathrm{AF}}(0)}_1\otimes\ket{\Psi_{\mathrm{AF}}(0)}_2  \nonumber \\
             &=&\left( \cos\theta \ket{e,1}+\sin\theta e^{-i\phi}\ket{g,2}\right)_1\otimes\left(\ket{g,2} \right)_2,
 \end{eqnarray}
 where $\ket{\Psi_{\mathrm{AF}}(0)}_i$ describes the $i$th atom-field subsystem and $\ket{n}$ ($n=1,2$) implies the number of photons in the nonperfect cavities. Notice that the initial atom-field which numbered 1(2) is entangled (separable). The Hamiltonian (\ref{eq1}) in the interaction picture for each atom-field cavity is as follows:
 \begin{eqnarray}\label{inthamilton}
 \hat{H}_\mathrm{int}=g \left( e^{i(\lambda-2\chi  \hat{a}^{\dagger}\hat{a})t} \hat{a}\hat{\sigma}^++\hat{a}^{\dagger} \hat{\sigma}^-e^{-i(\lambda-2\chi  \hat{a}^{\dagger}\hat{a})t}\right),
 \end{eqnarray}

 where
 \begin{eqnarray}\label{landa}
 \lambda&=&\Delta+i\frac{\kappa}{2}-i\frac{\varGamma}{2},\\
 \Delta&=&\omega-\nu,
 \end{eqnarray}
  where we assumed that the two subsystems are similar, i.e., $\omega_1=\omega_2\equiv \omega$, $\nu_1=\nu_2\equiv \nu$, $g_1=g_2\equiv g $, $\chi_1=\chi_2\equiv \chi$, $\kappa_1=\kappa_2=\kappa$ and $\varGamma_1=\varGamma_2=\varGamma$. The Hamiltonian (\ref{inthamilton}) leads to the following general (unnormalized) atom-field entangled states for each subsystem:
 \begin{subequations}\label{afstate}
    \begin{eqnarray}
       \ket{\varPsi_{\mathrm{AF}}(t)}_1&=& C_1(t)\ket{e,1}+C_2(t)\ket{g,2}, \\
       \ket{\varPsi_{\mathrm{AF}}(t)}_2&=& C_3(t)\ket{e,1}+C_4(t)\ket{g,2},
    \end{eqnarray}
 \end{subequations}
 in which
   \begin{subequations}\label{coff}
 \begin{eqnarray}
        C_1(t)&=& \left(\eta(t)\cos\theta-\frac{ig2\sqrt{2}}{\Omega}\sinh\left( \Omega t/2\right)e^{-i\phi}\sin\theta   \right)e^\frac{it\zeta}{2} ,  \label{A1} \\
        C_2(t)&=& \left(\eta'(t) e^{-i\phi}\sin\theta-\frac{ig2\sqrt{2}}{\Omega}\sinh\left( \Omega t/2\right)\cos\theta\right)e^{-\frac{it\zeta}{2}} ,  \label{A2} \\
        C_3(t)&=&-\frac{ig2\sqrt{2}}{\Omega}\sinh\left( \Omega t/2\right)e^\frac{it\zeta}{2},  \label{A3} \\
        C_4(t)&=& \eta'(t)e^{-\frac{it\zeta}{2}}, \label{A4}
 \end{eqnarray}
 \end{subequations}

  where as is found in (\ref{initialstate}) the parameter $\theta$  determines the probability amplitudes of the initial atom-field state and
 \begin{subequations}\label{variables}
 \begin{eqnarray}
          \eta(t)&=&-i\frac{\zeta}{\Omega}\sinh\left( \Omega t/2\right)+\cosh\left(\Omega t/2 \right),  \\
          \eta'(t)&=&i\frac{\zeta}{\Omega}\sinh\left( \Omega t/2\right)+\cosh\left(\Omega t/2 \right),  \\
          \Omega&=&\sqrt{-\zeta^2-8g^2},  \\
          \zeta &=&\lambda-2\chi.
 \end{eqnarray}
 \end{subequations}
 A rigorous attention to our above derivations and definitions reveals that $\Omega$ is not the usual Rabi frequency. By a glance at Eqs. (\ref{inthamilton}) and (\ref{landa}) it is evident that when $\kappa=\varGamma$ one arrives at $\lambda=\Delta$ and our system behaves as an ideal system with no dissipation. Therefore, seemingly this is a useful and interesting condition that allows one to get rid of the dissipation. Notice that due to the non-hermiticity of Hamiltonian (\ref{eq1}) we should normalize the atom-field states (\ref{afstate}) as follows:
 \begin{subequations}\label{aft}
    \begin{eqnarray}
       \ket{\Psi_{\mathrm{AF}}(t)}_1&=& A_1(t)\ket{e,1}+A_2(t)\ket{g,2}, \label{aft1} \\
       \ket{\Psi_{\mathrm{AF}}(t)}_2&=& A_3(t)\ket{e,1}+A_4(t)\ket{g,2}, \label{aft2}
    \end{eqnarray}
 \end{subequations}
 where the coefficients $A_i(t)$ ($i=1,\ldots,4$) can be easily calculated as below:
   \begin{eqnarray}\label{Ncoef}
       A_{1(2)}(t)&=&\frac{C_{1(2)}(t)}{\sqrt{|C_1(t)|^2+|C_2(t)|^2}},  \\ \nonumber
       A_{3(4)}(t)&=&\frac{C_{3(4)}(t)}{\sqrt{|C_3(t)|^2+|C_4(t)|^2}}.\nonumber
    \end{eqnarray}
     As is observed, there is no direct interaction among these two subsystems  $(\mathrm{AF})_i$ ($i=1,2$), therefore their states are expected to remain separable with the density operator:
          \begin{equation}\label{statet}
          \hat{\rho}(t)= \ket{\Psi(t)}\bra{\Psi(t)},
        \end{equation}
  in which $\ket{\Psi(t)}=\ket{\Psi_{\mathrm{AF}}(t)}_1\otimes\ket{\Psi_{\mathrm{AF}}(t)}_2$. It is clearly seen from  Eqs. (\ref{aft1}) and (\ref{aft2}) that  in each cavity the states of atom-field are entangled.\\Now, in the line of the goals of paper, it is quite reasonable to search for a strategy to exchange the entanglement between atom-field in each cavity into atom-atom (and/or field-field between the two cavity fields) for the sake of quantum information processing tasks. In this regard, one could think of creating entanglement between the two atoms by performing the BSM onto the field modes leaving the cavities (see Fig. \ref{fig:model}). Mathematically speaking, this task can be done by projection $\ket{\Psi(t)}$ onto one of the suitable Bell states. Among different types of resources for linear optical quantum swapping implementations \cite{Lee2013}, the two-photon pairs has been put forward to be an efficient resource for our purpose \cite{Lee2013}. Taking a look at Eq. (\ref{aft}) reveals that there are only one and two photons in the field states. Henceforth, it is quite reasonable to consider the following maximally entangled state\footnote{The properties of Bell state can be seen in \cite{gisin1998}. Interestingly, for two atomic system states (two qubits) \cite{Chen2007} the Bell state is defined as $\ket{\Psi^\pm}=\frac{1}{\sqrt{2}}\left( \ket{e,g}\pm\ket{g,e}\right)$,  $\ket{\Phi^\pm}=\frac{1}{\sqrt{2}}\left( \ket{e,e}\pm\ket{g,g}\right)$, and even the Bell states for the compound atom-field state  have been  defined as $\ket{\Phi^\pm}=\frac{1}{\sqrt{2}}\left( \ket{e,1}\pm\ket{g,0}\right)$, $\ket{\Psi^\pm}=\frac{1}{\sqrt{2}}\left( \ket{e,0}\pm\ket{g,1}\right)$ \cite{Chen2007}. In addition, considering the two possible polarizations $\ket{V}$ and $\ket{H}$ some other Bell type states have been constructed \cite{Sheng2010}.}
       \begin{equation}
       \label{bellfield}
 \ket{\Psi^-}_\mathrm{F}=\frac{1}{\sqrt{2}}\left( \ket{1,2}-\ket{2,1} \right),
       \end{equation}
 in order to construct the projection operator, as $\hat{P}^-_\mathrm{F}=\ket{\Psi^-}{}_\mathrm{FF}\bra{{}\Psi^-}$.  Consequently, the operation of this projection operator on above $\ket{\Psi(t)}$ leaves the field states and establishes the following (unnormalized) atom-atom entangled state:
 \begin{equation}\label{AAunnorm}
      \ket{\Psi_\mathrm{AA}(t)}= {}_\mathrm{F}\left\langle \Psi^- |\Psi(t)\right\rangle  =\frac{1}{\sqrt{2}}[A_1(t)A_4(t)\ket{e,g}-A_2(t)A_3(t)\ket{g,e}].
 \end{equation}
 Normalization of the above state arrives one at the following atomic reduced density operator:
       \begin{equation}\label{densityoperator}
        \hat{\rho}_\mathrm{AA}(t)=\frac{\ket{\Psi_\mathrm{AA}(t)}\bra{\Psi_\mathrm{AA}(t)}}{\mathrm{Tr}\left(\ket{\Psi_\mathrm{AA}(t)}\bra{\Psi_\mathrm{AA}(t)} \right) },
       \end{equation}
         in which
       \begin{equation}\label{eq8}
   \Tr\left(\ket{\Psi_\mathrm{AA}(t)}\bra{\Psi_\mathrm{AA}(t)} \right)=\frac{1}{2}\left( \left| A_1(t)\right| ^2\left| A_4(t)\right|^2+\left| A_2(t)\right|^2\left| A_3(t)\right|^2\right) =N(t).
       \end{equation}
 In order to quantify the amount of created entanglement between the two atoms, we use the concurrence \cite{Wootters1998}  which has been defined as
        \begin{equation}\label{con}
         {\cal C}(t)=\mathrm{max}\{0,\sqrt{\lambda_1}-\sqrt{\lambda_2}-\sqrt{\lambda_3}-\sqrt{\lambda_4}\},
        \end{equation}
 where $\lambda_i$ ($i=1,\ldots,4$) are the eigenvalues (in decreasing order) of the Hermitian matrix
 $\hat{\rho}_\mathrm{AA}\left(\sigma_1^y\otimes\sigma_2^y\hat{\rho}_\mathrm{AA}^{*}\sigma_1^y\otimes\sigma_2^y\right)$ with $\hat{\rho}_\mathrm{AA}^*$ as the complex conjugate of $\hat{\rho}_\mathrm{AA}$ and $\sigma_k^y:=i(\sigma_k-\sigma_k^\dag)$.  The concurrence varies between 0 (completely separable) and 1 (maximally entangled). With the help of (\ref{densityoperator}) and (\ref{con}), the explicit form of the concurrence can be obtained as below:
        \begin{equation}\label{cont}
         {\cal C}(t)=\frac{\left| A_1(t)\right| \left|A_2(t)\right| \left| A_3(t)\right| \left| A_4(t)\right| }{N(t)}.
        \end{equation}
 Having obtained the explicit form of the atom-atom entangled state, we are also able to find the occupation probabilities of the atom-atom states. Furthermore, a glance at Eq. (\ref{AAunnorm}) reveals that it is only possible to have transitions from the state $\ket{e,g}$ into $\ket{g,e}$. Therefore, only these two occupation probabilities are available:
 \begin{subequations}\label{eqp}
 \begin{eqnarray}
 P_1(t)&=& \bra{e,g}\hat{\rho}_\mathrm{AA}(t)\ket{e,g}=\frac{\left| A_1(t)\right|^2\left| A_4(t)\right|^2}{2N(t)}, \label{eqp1} \\
 P_2(t)&=& \bra{g,e}\hat{\rho}_\mathrm{AA}(t)\ket{g,e}=\frac{\left| A_2(t)\right|^2\left| A_3(t)\right|^2}{2N(t)}.  \label{eqp2}
 \end{eqnarray}
 \end{subequations}
 Clearly, according to (\ref{eq8}), (\ref{eqp1}) and (\ref{eqp2}),  $P_1(t)+P_2(t)=1$ at any time $t$.

\section{Results and discussion} \label{sec.results}
\begin{figure}[ht]
    \centering
 \includegraphics[width=0.5\textwidth]{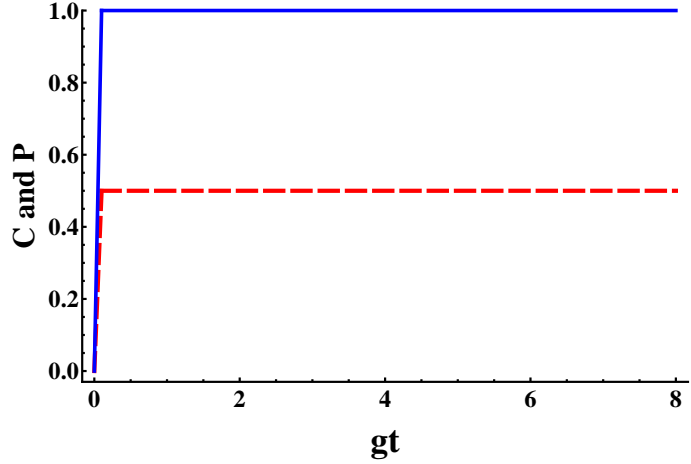}
    \caption{\label{fig:tpi} The concurrence (solid blue line) and occupation probabilities (dashed red line) as functions of scaled time $gt$ for $\theta=(2n+1)\pi/2$ and any values of $\chi$, $\Delta$, $\kappa$ and $\varGamma$.}
   \end{figure}

\begin{figure}[ht]
    \centering
 \includegraphics[width=0.6\textwidth]{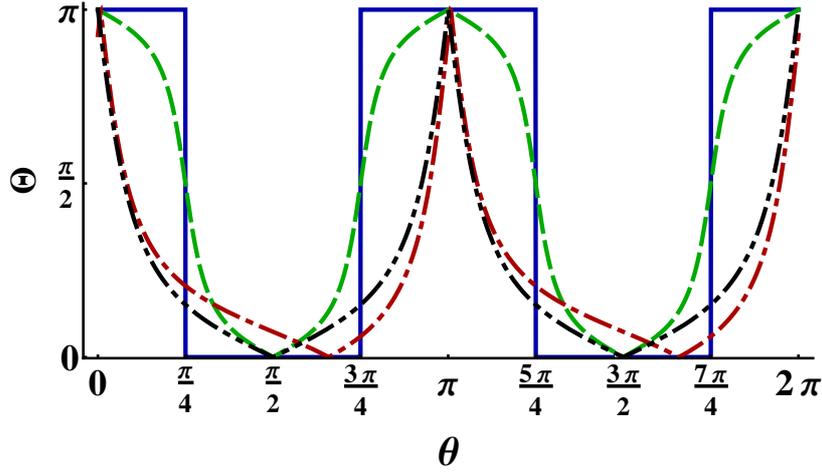}
    \caption{\label{fig.Theta} The phase $\Theta$ as function of $\theta$ at time intervals obtained in Eq. (\ref{Tnteta0}) with $\phi=\pi/2$ for $\Delta/g=\chi/g=0$ (solid blue line), $\Delta/g=0$, $\chi/g=0.4$ (dashed green line), $\Delta/g=7$, $\chi/g=0$ (dot-dashed red line) and $\Delta/g=7$, $\chi/g=0.4$ (dot-dot-dashed black line) for $\kappa=\varGamma$.}
   \end{figure}

In this section, we shall examine the role of initial state, detuning parameter and Kerr medium on the dynamics of atom-atom swapped entanglement in absence and presence of dissipation. Before considering the time evolution of entanglement, let us check the situations in which the concurrence can reach its maximum value. Taking a look at Eq. (\ref{AAunnorm})  reveals that, whenever $P_1(t)=P_2(t)=0.5$ (from Eq. (\ref{eqp})), the atom-atom state is projected into the following maximally entangled state:
  \begin{eqnarray}\label{Bellstate}
  \ket{\Psi_\mathrm{AA}}=\frac{1}{\sqrt{2}}(\ket{e,g}-e^{i \Theta(t)}\ket{g,e}),
  \end{eqnarray}
in which
\begin{equation}
\Theta(t)\equiv\phi_{14}(t)-\phi_{23}(t),
\end{equation}
where we have defined $\phi_{ij}\equiv\phi_{i}+\phi_{j}$ and $A_{i}=\left|A_{i} \right|e^{-i\phi_{i}}$ ($i,j=1,...,4$). It should be noticed that this is not the only condition in which the maximum value of concurrence is occurred.  According to the above-mentioned condition and taking into account the Eqs. (\ref{coff}), (\ref{aft}), (\ref{eqp1}) and (\ref{eqp2})  it is straightforward to show that  whenever the general condition $\cos\theta=0$ is fulfilled, the concurrence reaches its maximum value. Going back to this condition, we simply have $\theta=(2n+1)\pi/2$, which means that the initial state of system is $\ket{g,2}\otimes\ket{g,2}$ (up to a global phase $\phi$).  In this case, before the occurrence of any interaction between the atom-field in both cavities, the result of BSM would be zero. Therefore, there is no swapped entanglement between atoms at the beginning of the interaction. The surprising aspect here is that, soon after starting the interaction, the concurrence suddenly jumps from zero to its  maximum value (i.e., 1), and this maximum remains fixed as time goes on, independent of the detuning parameter, the Kerr effect and dissipation parameters. The same behaviour is expected for the occupation probabilities (except the maximum value's which is 0.5). It is also interesting to note that with this condition (i.e., $\theta=(2n+1)\pi/2$), the phase $\Theta$ in Eq. (\ref{Bellstate}) is always zero, which means that the state of atom-atom has been projected into the following Bell state up to an irrelevant global phase (see the subscript in page 4):
  \begin{eqnarray}\label{Bellstate1}
  \ket{\Psi^-}=\frac{1}{\sqrt{2}}(\ket{e,g}-\ket{g,e}).
  \end{eqnarray}
 Therefore, the concurrence (occupation probabilities) takes the form ${\cal C}(t)=1-\delta(t)$ ($P_1= P_2=0.5[1-\delta(t)]$), in which $\delta$ is the usual Dirac delta function. These are shown in  Fig. \ref{fig:tpi}.\\
 By the assumption $\kappa=\varGamma$, we obtain the second condition that concurrence being equal to 1 as follows:
 \begin{equation}
  \left|  \eta(t)\right|  ^4-\frac{64g^4}{\Omega_0^4}\sinh^4\left( \Omega_0 t/2\right)=0.
 \end{equation}
   One can straightforwardly obtain the following times $T_n$ at which the concurrence is maximum:
 \begin{equation}\label{Tnteta0}
 T_n=\frac{2 n\pi}{\Omega_0}+\frac{2}{\Omega_0}\tan^{-1}\left( \frac{\Omega_0}{\sqrt{8g^2-(\Delta-2\chi)^2}} \right),
 \end{equation}
 where
  \begin{equation}\label{omega0}
 \Omega_0=\sqrt{(\Delta-2\chi)^2+8g^2},
  \end{equation}
 in which $n$ is an integer. This result is valid for any value of $\theta\neq(2n+1)\pi/2$. At these instants of time, the state of atom-atom after BSM is projected into the state (\ref{Bellstate}). The phase $\Theta$ determines whether this state belongs to the family of Bell states. In Fig. \ref{fig.Theta} we have plotted the phase $\Theta$ as a function of $\theta$ at time intervals obtained in Eq. (\ref{Tnteta0}) for some values of detuning and  Kerr medium parameter. It is evident that, in the absence of detuning parameter and Kerr medium, phase $\Theta$ has a step behaviour versus the angle $\theta$. Therefore, as $\theta$ increases in this case, the state of  atom-atom  interpolates between $ \ket{\Psi^-}$ in Eq. (\ref{Bellstate1}) and
 \begin{equation}\label{Bellp}
 \ket{\Psi^+}=\frac{1}{\sqrt{2}}(\ket{e,g}+\ket{g,e}).
 \end{equation}
 For nonzero values of detuning and Kerr medium parameters, Bell state $\ket{\Psi^+}$ is achieved for $\theta=n\pi$ ($n=0,1,2,...$).

\begin{figure}[ht]
\centering
\subfigure[\label{fig.n5} \ $\frac{\Delta}{g}=\frac{\chi}{g}=0$, $\frac{\kappa}{g}=\frac{\varGamma}{g}$]{\includegraphics[width=0.35\textwidth]{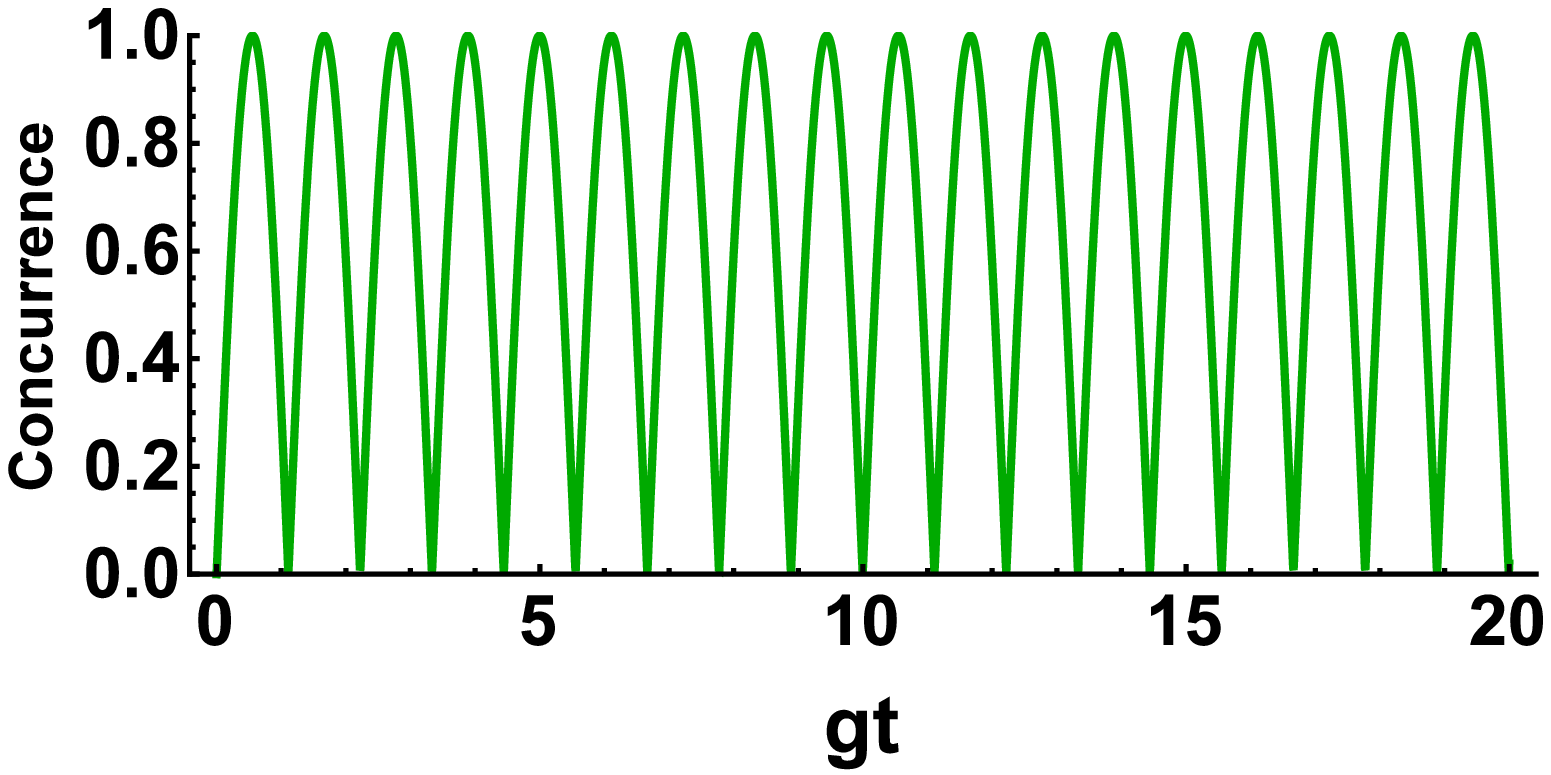}}
\hspace{0.05\textwidth}
\subfigure[\label{fig.nn9} \ $\frac{\Delta}{g}=10$, $\frac{\chi}{g}=0$, $\frac{\kappa}{g}=\frac{\varGamma}{g}$]{\includegraphics[width=0.35\textwidth]{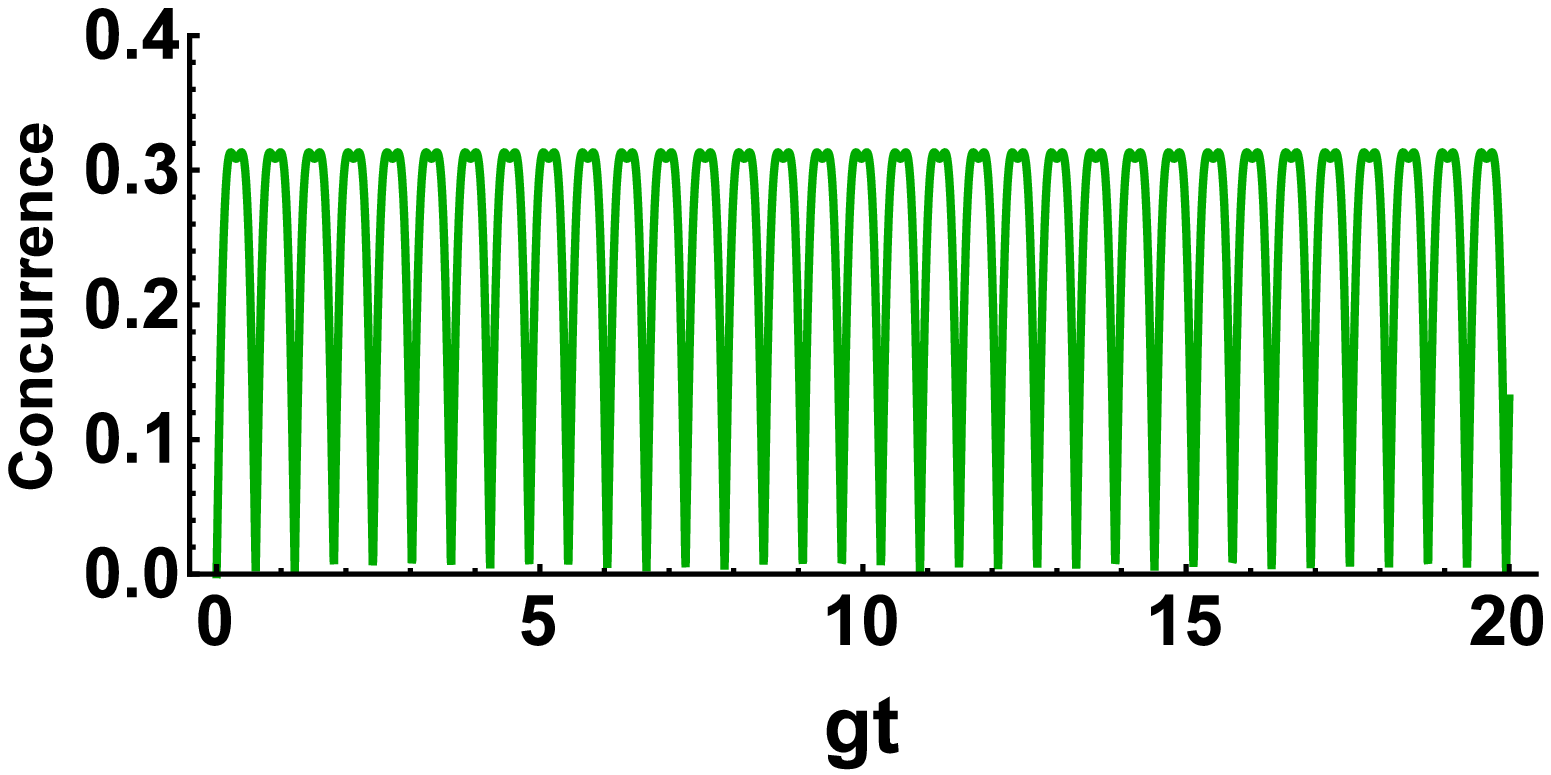}}
\hspace{0.05\textwidth}
\subfigure[\label{fig.10} \ $\frac{\Delta}{g}=10$, $\frac{\chi}{g}=0.4$, $\frac{\kappa}{g}=\frac{\varGamma}{g}$]{\includegraphics[width=0.35\textwidth]{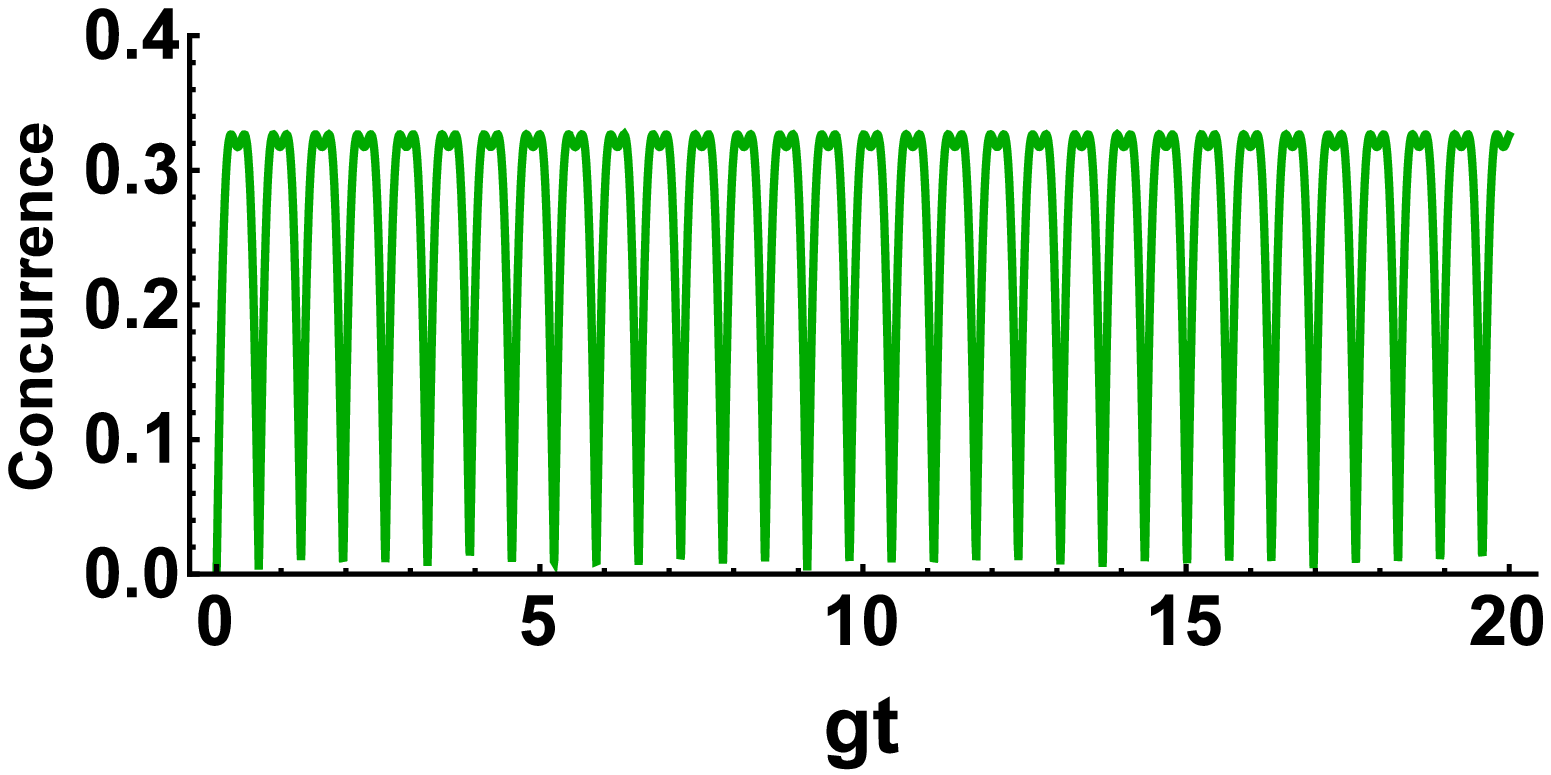}}
\hspace{0.05\textwidth}
\subfigure[\label{fig.11} \ $\frac{\Delta}{g}=0$, $\frac{\chi}{g}=0.4$, $\frac{\kappa}{g}=\frac{\varGamma}{g}$]{\includegraphics[width=0.35\textwidth]{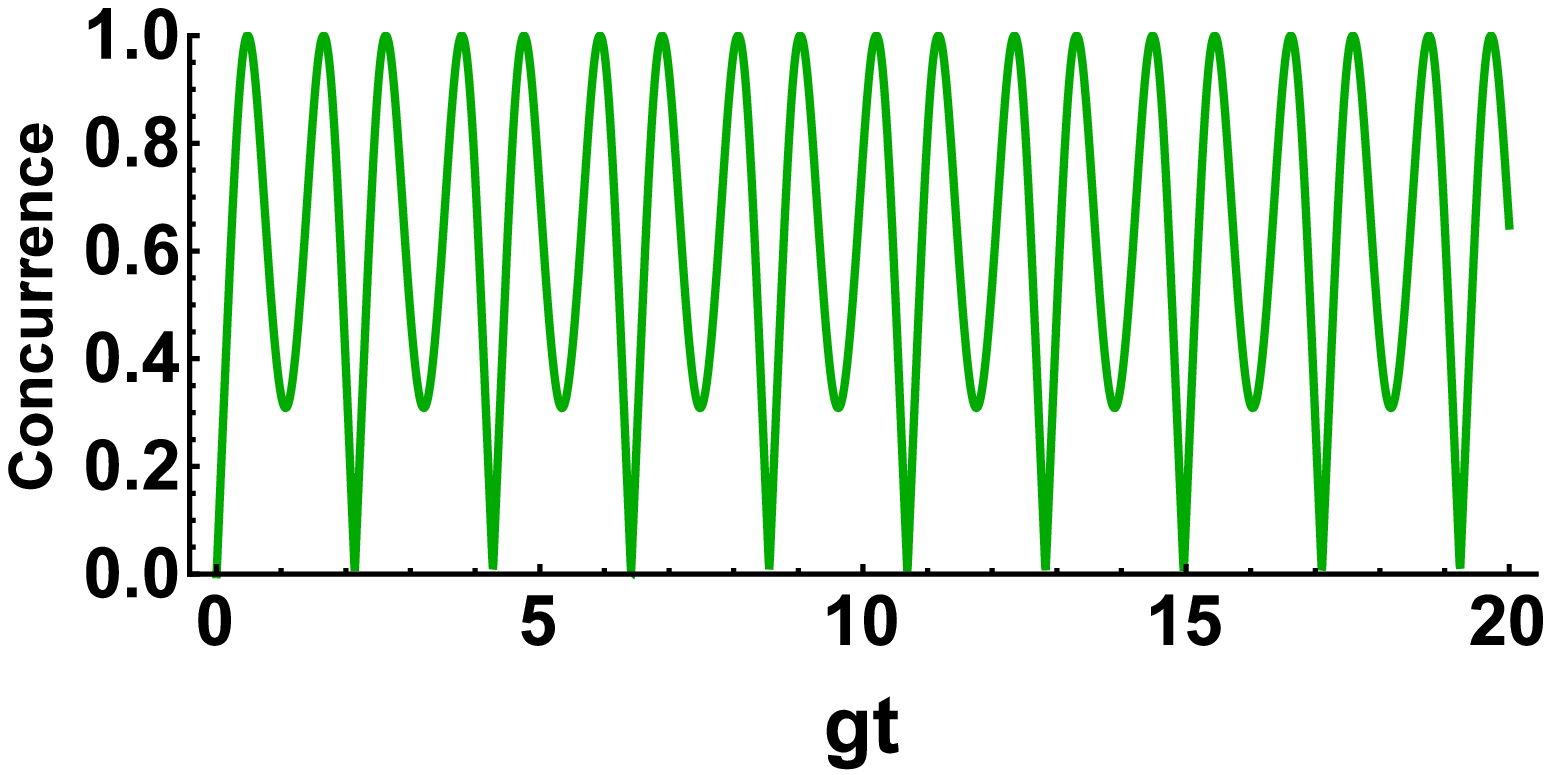}}
\hspace{0.05\textwidth}
\subfigure[\label{fig.nn} \ $\frac{\Delta}{g}=0$, $\frac{\chi}{g}=1$, $\frac{\kappa}{g}=\frac{\varGamma}{g}$]{\includegraphics[width=0.35\textwidth]{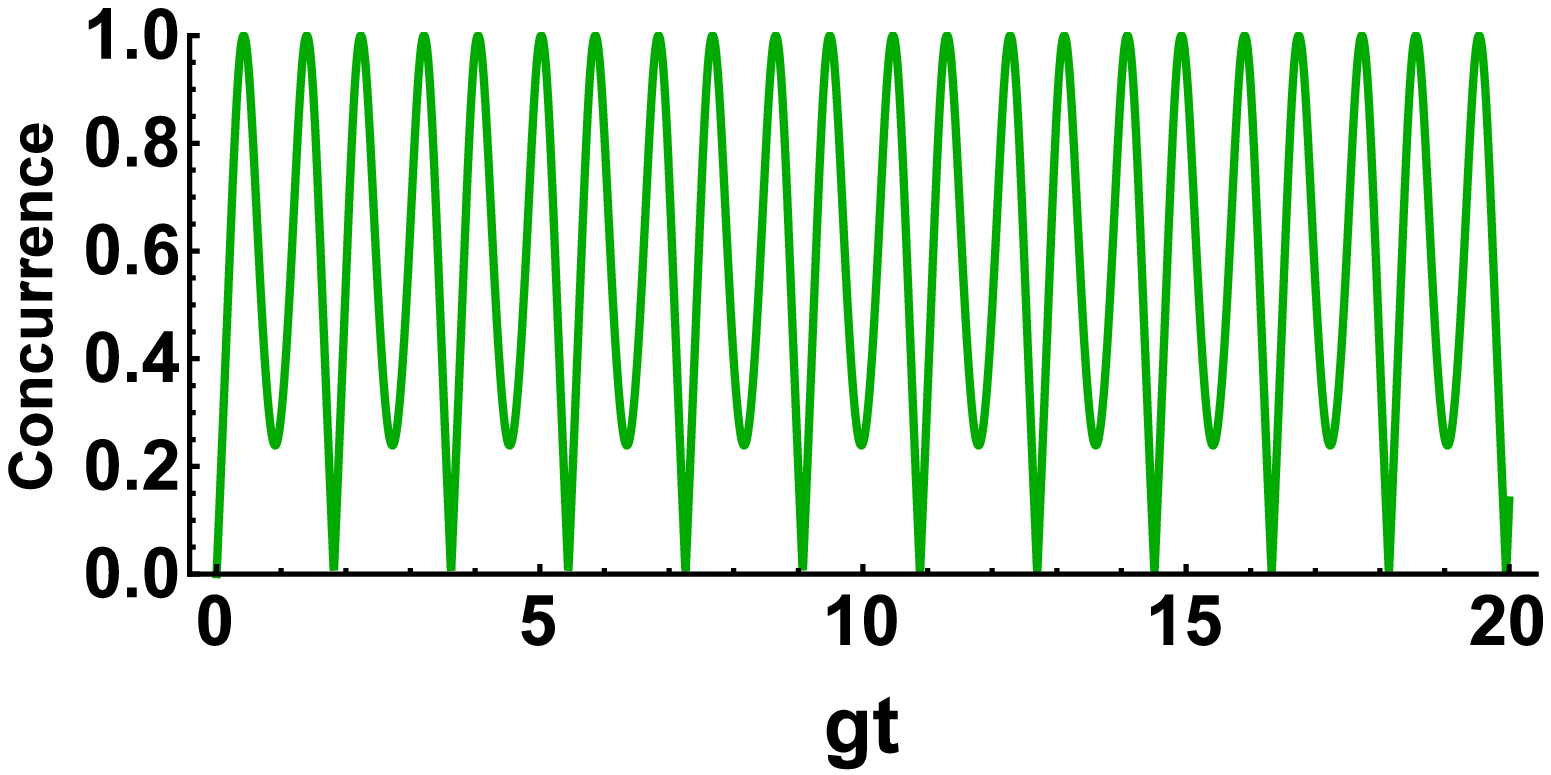}}
\hspace{0.05\textwidth}
\subfigure[\label{fig.n6} \ $\frac{\Delta}{g}=10$, $\frac{\chi}{g}=0$, $\frac{\kappa}{g}=0.1$, $\frac{\varGamma}{g}=0.3$]{\includegraphics[width=0.35\textwidth]{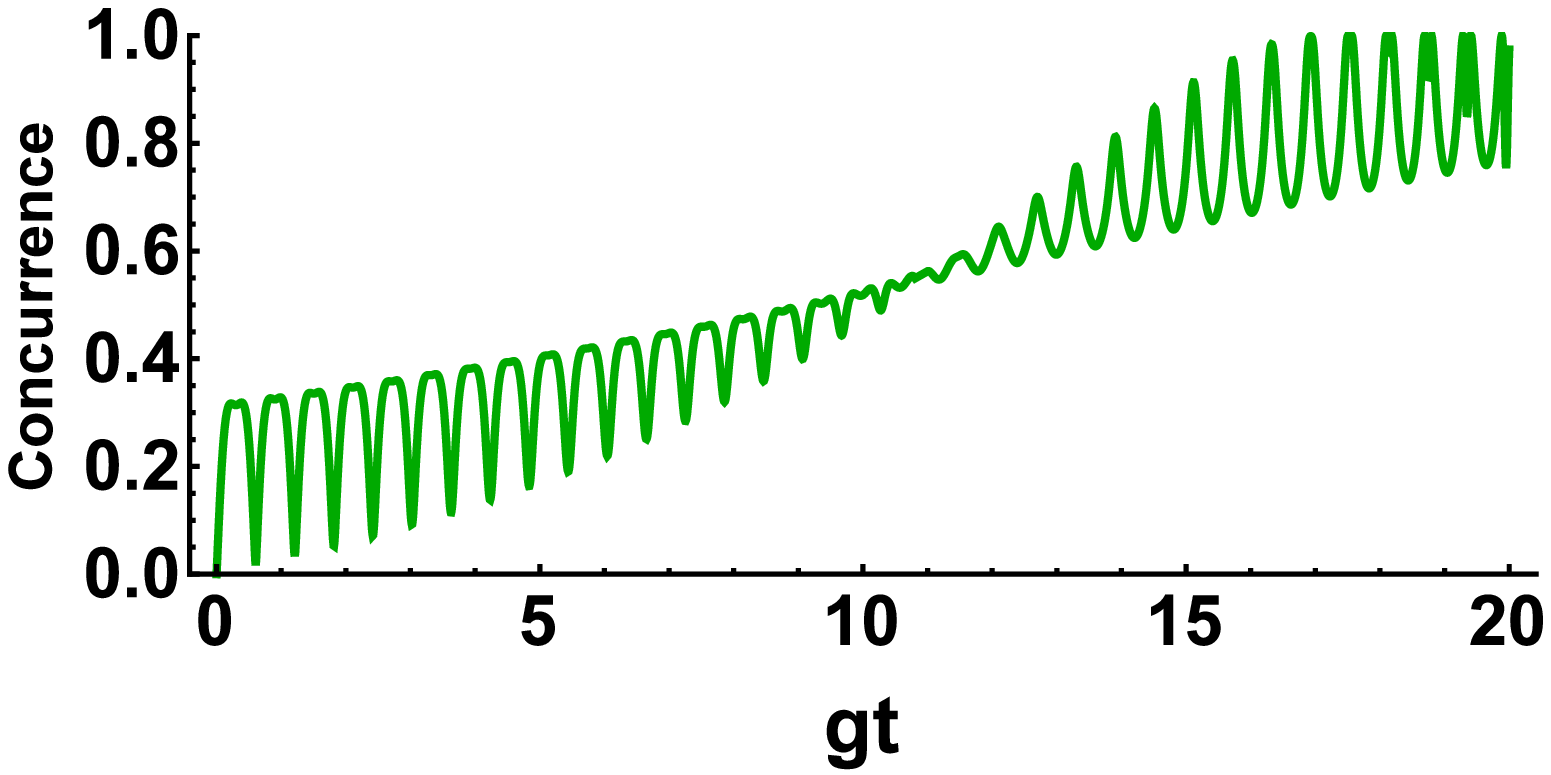}}
\hspace{0.05\textwidth}
\subfigure[\label{fig.13} \ $\frac{\Delta}{g}=10$, $\frac{\chi}{g}=0.4$, $\frac{\kappa}{g}=2$, $\frac{\varGamma}{g}=3$]{\includegraphics[width=0.35\textwidth]{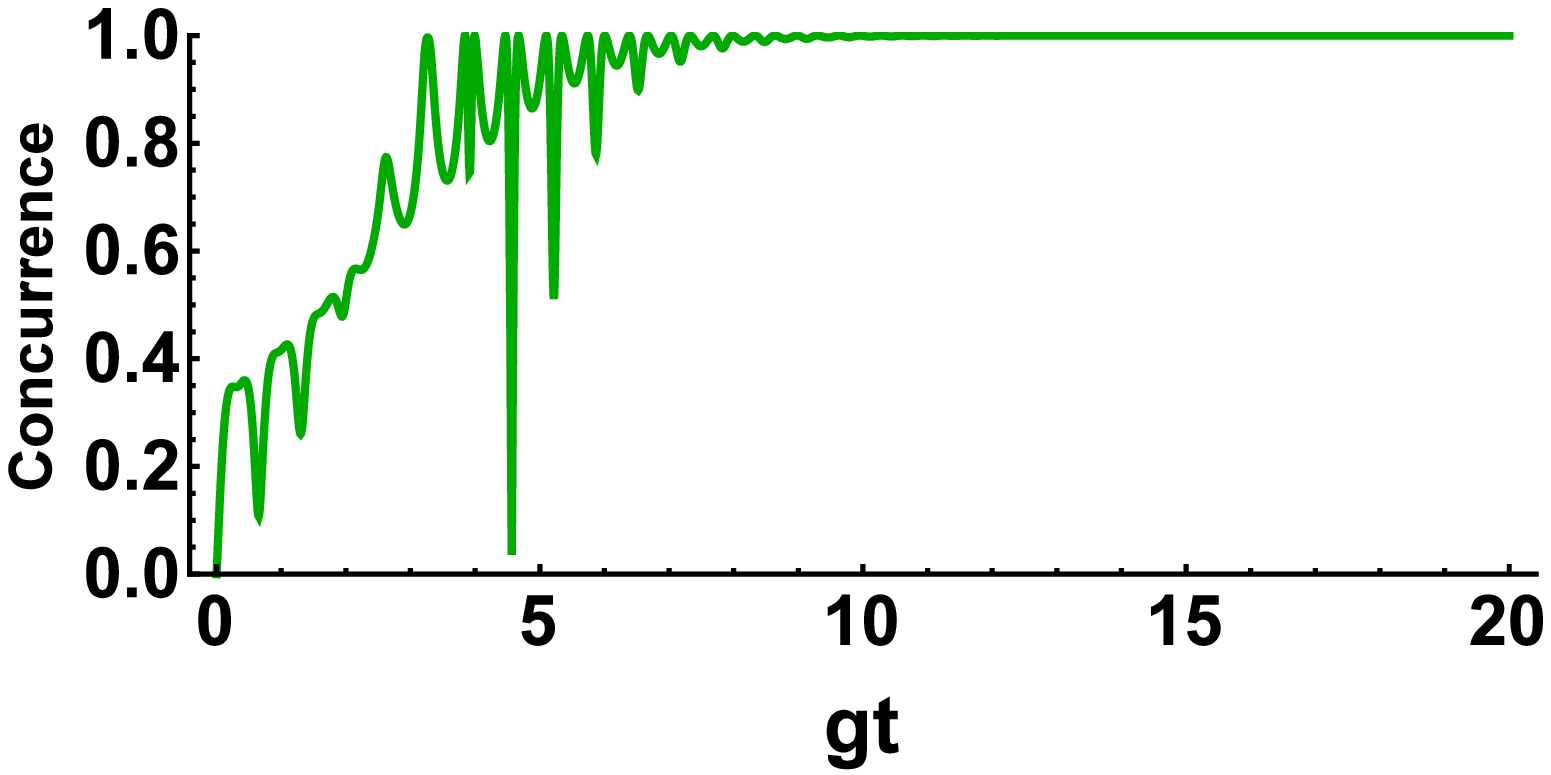}}
\caption{\label{fig.tpi4} The time evolution of concurrence after entanglement swapping as function of the scaled time $gt$ when $\theta=\frac{\pi}{4}$, $\phi=0$.}
\end{figure}

\begin{figure}[ht]
\centering
\subfigure[\label{fig.1} \ $\frac{\Delta}{g}=\frac{\chi}{g}=0$, $\frac{\kappa}{g}=\frac{\varGamma}{g}$]{\includegraphics[width=0.35\textwidth]{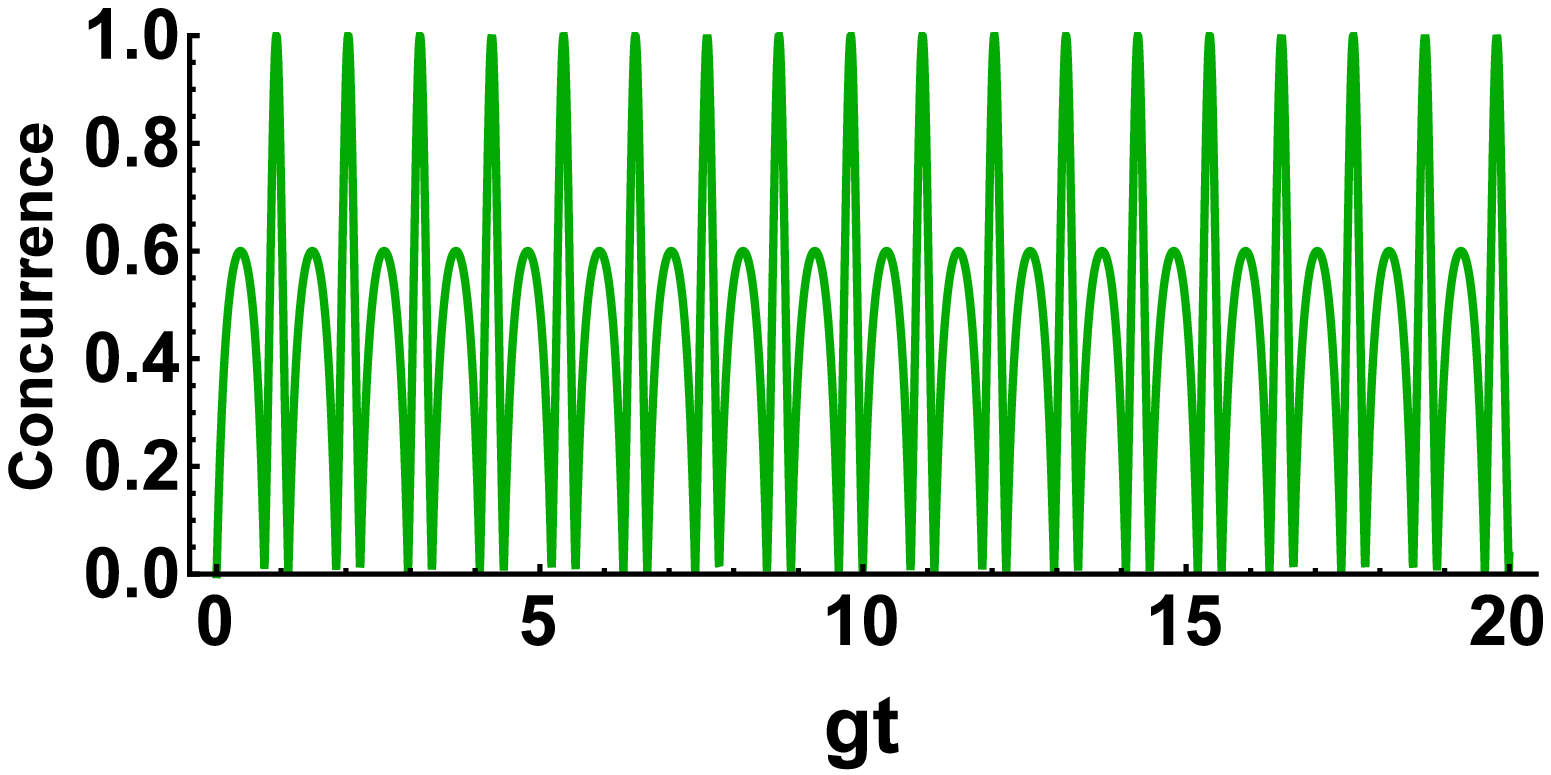}}
\hspace{0.05\textwidth}
\subfigure[\label{fig.5} \ $\frac{\Delta}{g}=10$, $\frac{\chi}{g}=0$, $\frac{\kappa}{g}=\frac{\varGamma}{g}$]{\includegraphics[width=0.35\textwidth]{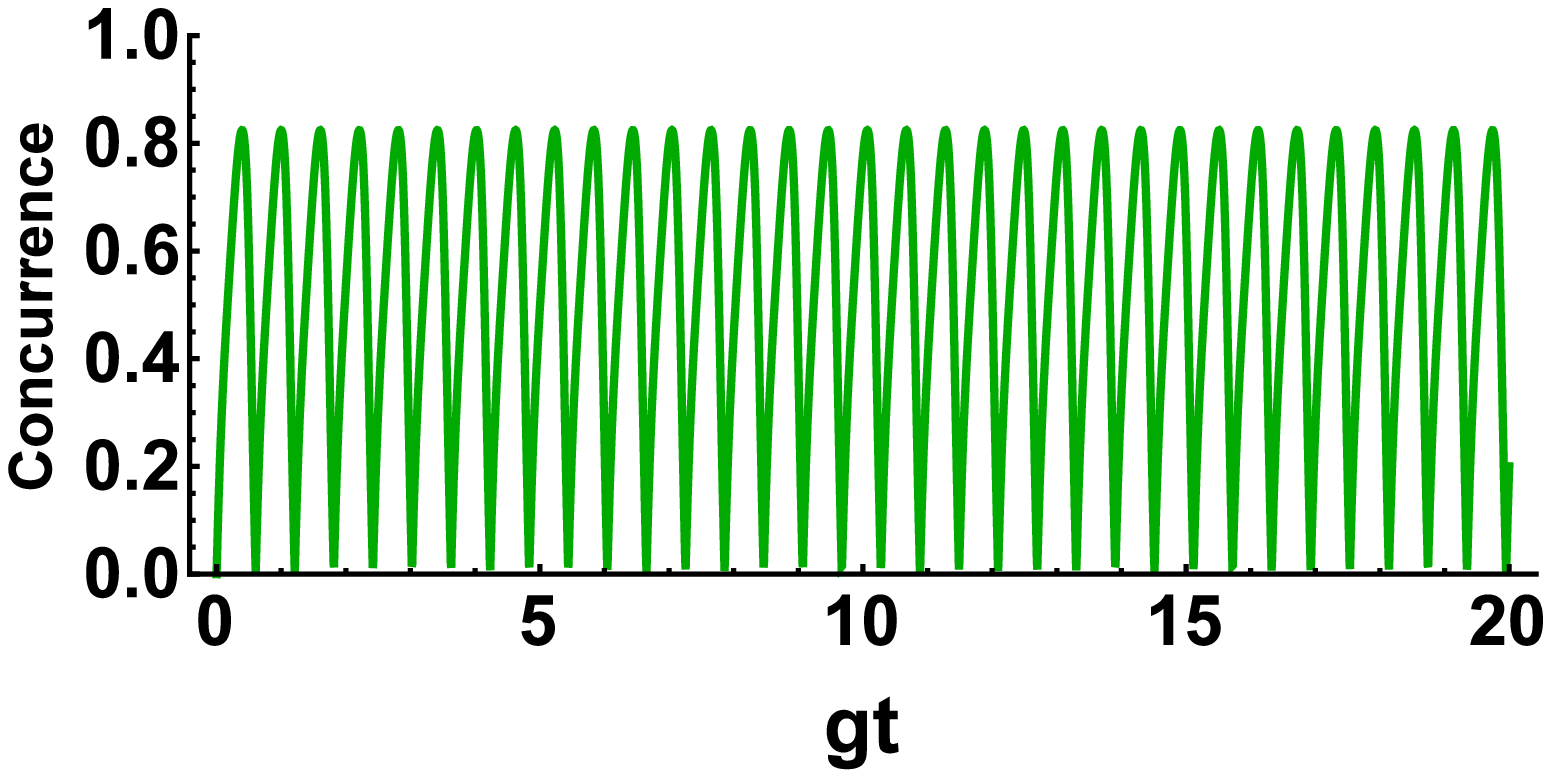}}
\hspace{0.05\textwidth}
\subfigure[\label{fig.nn5} \ $\frac{\Delta}{g}=7$, $\frac{\chi}{g}=0.7$, $\frac{\kappa}{g}=\frac{\varGamma}{g}$]{\includegraphics[width=0.35\textwidth]{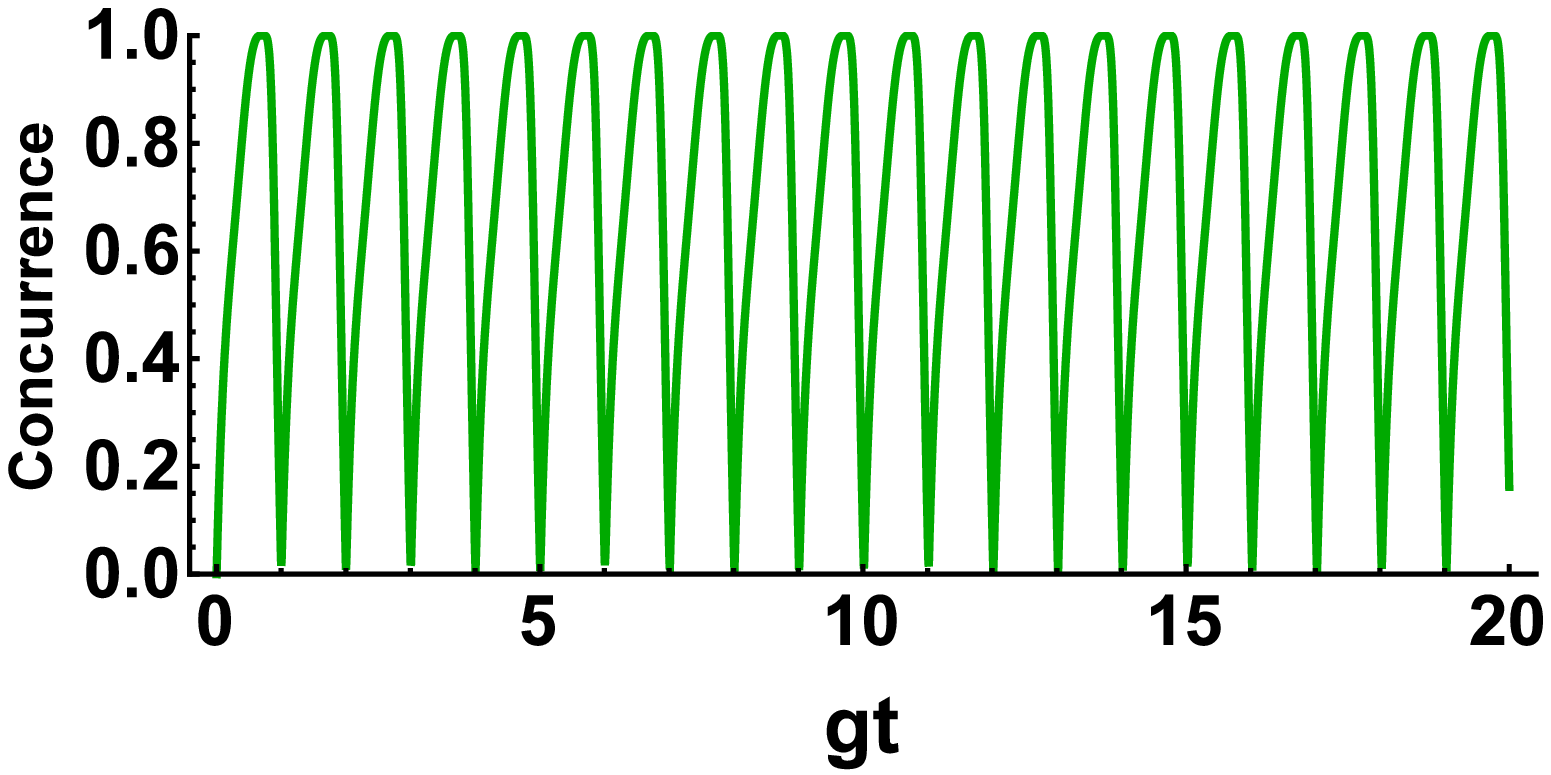}}
\hspace{0.05\textwidth}
\subfigure[\label{fig.6} \ $\frac{\Delta}{g}=0$, $\frac{\chi}{g}=0.7$, $\frac{\kappa}{g}=\frac{\varGamma}{g}$]{\includegraphics[width=0.35\textwidth]{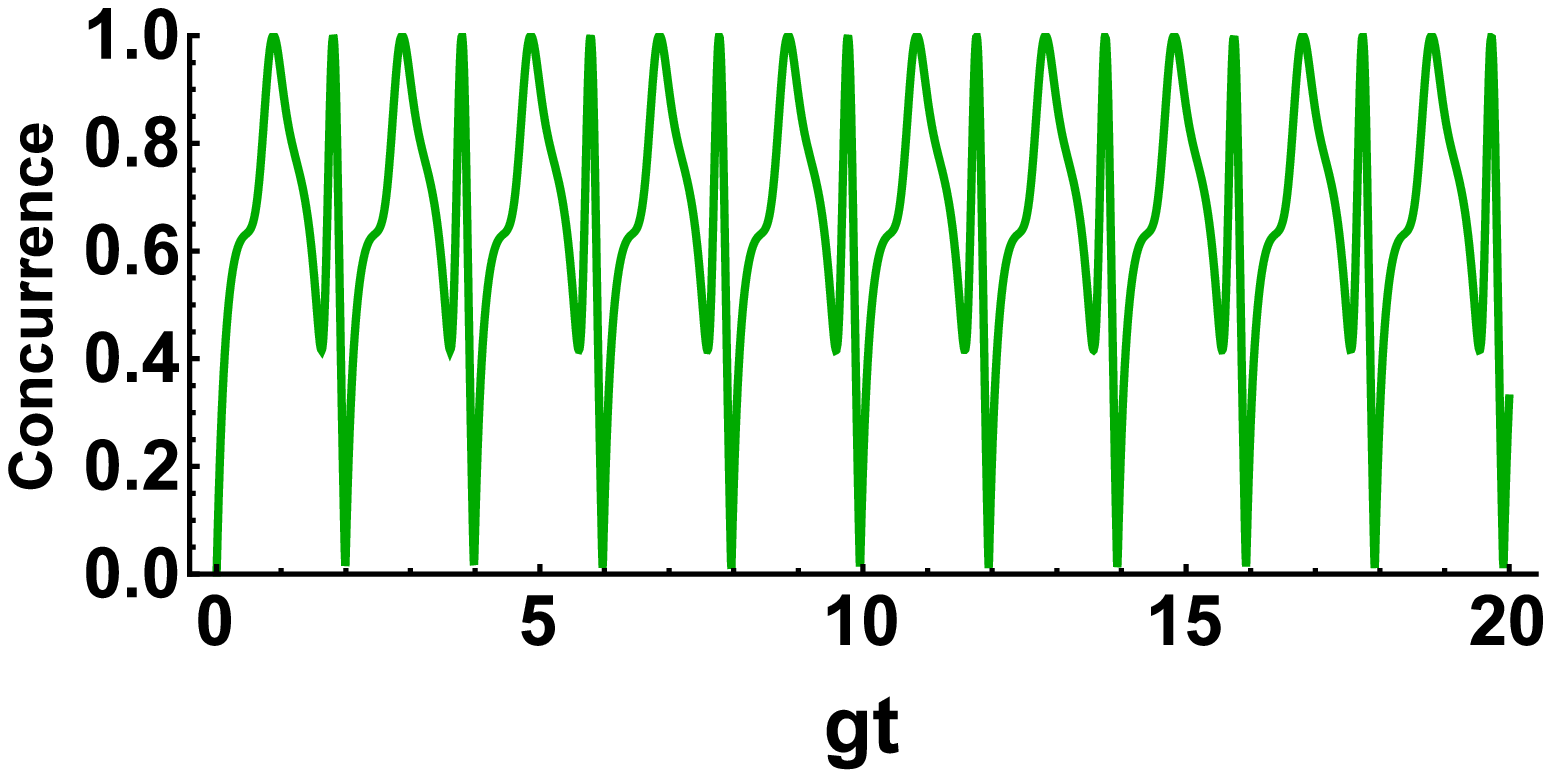}}
\hspace{0.05\textwidth}
\subfigure[\label{fig.ll} \ $\frac{\Delta}{g}=7$, $\frac{\chi}{g}=0$, $\frac{\kappa}{g}=0.1, \frac{\varGamma}{g}=0.3$]{\includegraphics[width=0.35\textwidth]{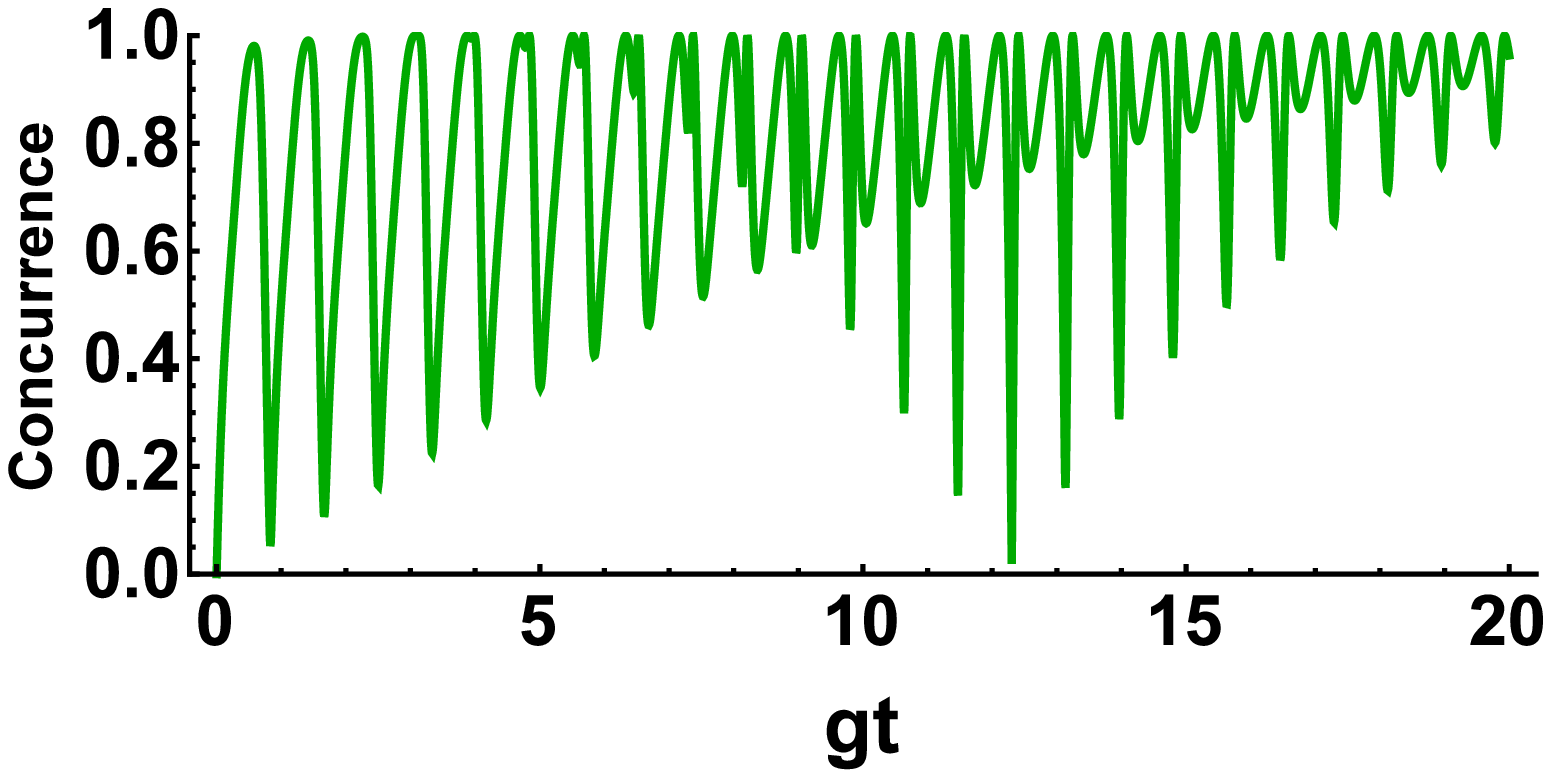}}
\hspace{0.05\textwidth}
\subfigure[\label{fig.2} \ $\frac{\Delta}{g}=7$, $\frac{\chi}{g}=0.7$, $\frac{\kappa}{g}=2$, $\frac{\varGamma}{g}=3$]{\includegraphics[width=0.35\textwidth]{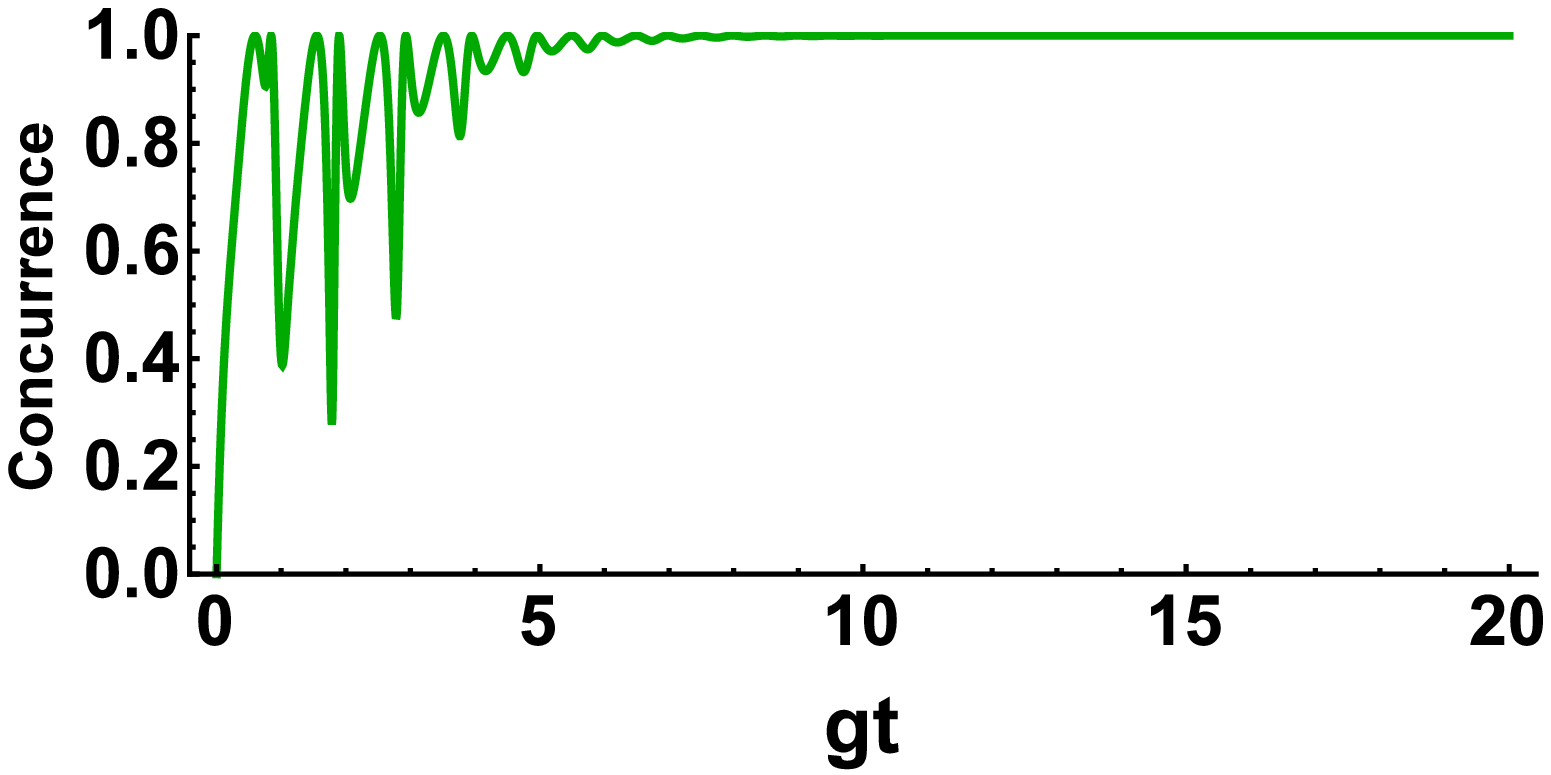}}
\caption{\label{fig.conp} The time evolution of concurrence after entanglement swapping as function of the scaled time $gt$ when $\theta=\frac{\pi}{3}$, $\phi=\frac{3\pi}{2}$.}
\end{figure}

\begin{figure}[ht]
\centering
\subfigure[\label{fig.p1} \ $\frac{\Delta}{g}=10$, $\frac{\chi}{g}=0$, $\frac{\kappa}{g}=0.1$, $\frac{\varGamma}{g}=0.3$]{\includegraphics[width=0.35\textwidth]{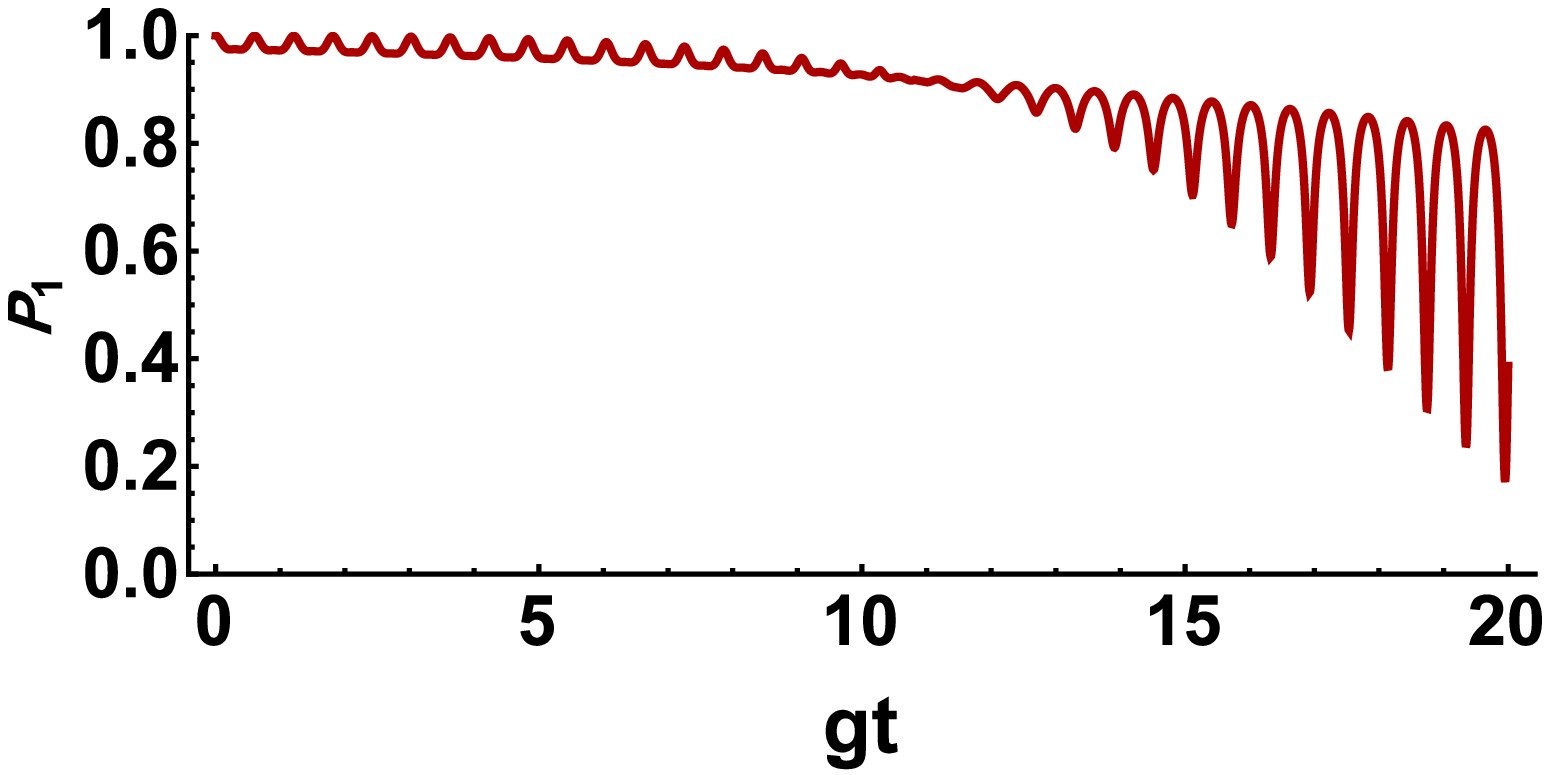}}
\hspace{0.05\textwidth}
\subfigure[\label{fig.1313} \ $\frac{\Delta}{g}=10$, $\frac{\chi}{g}=0.4$, $\frac{\kappa}{g}=2$, $\frac{\varGamma}{g}=3$]{\includegraphics[width=0.35\textwidth]{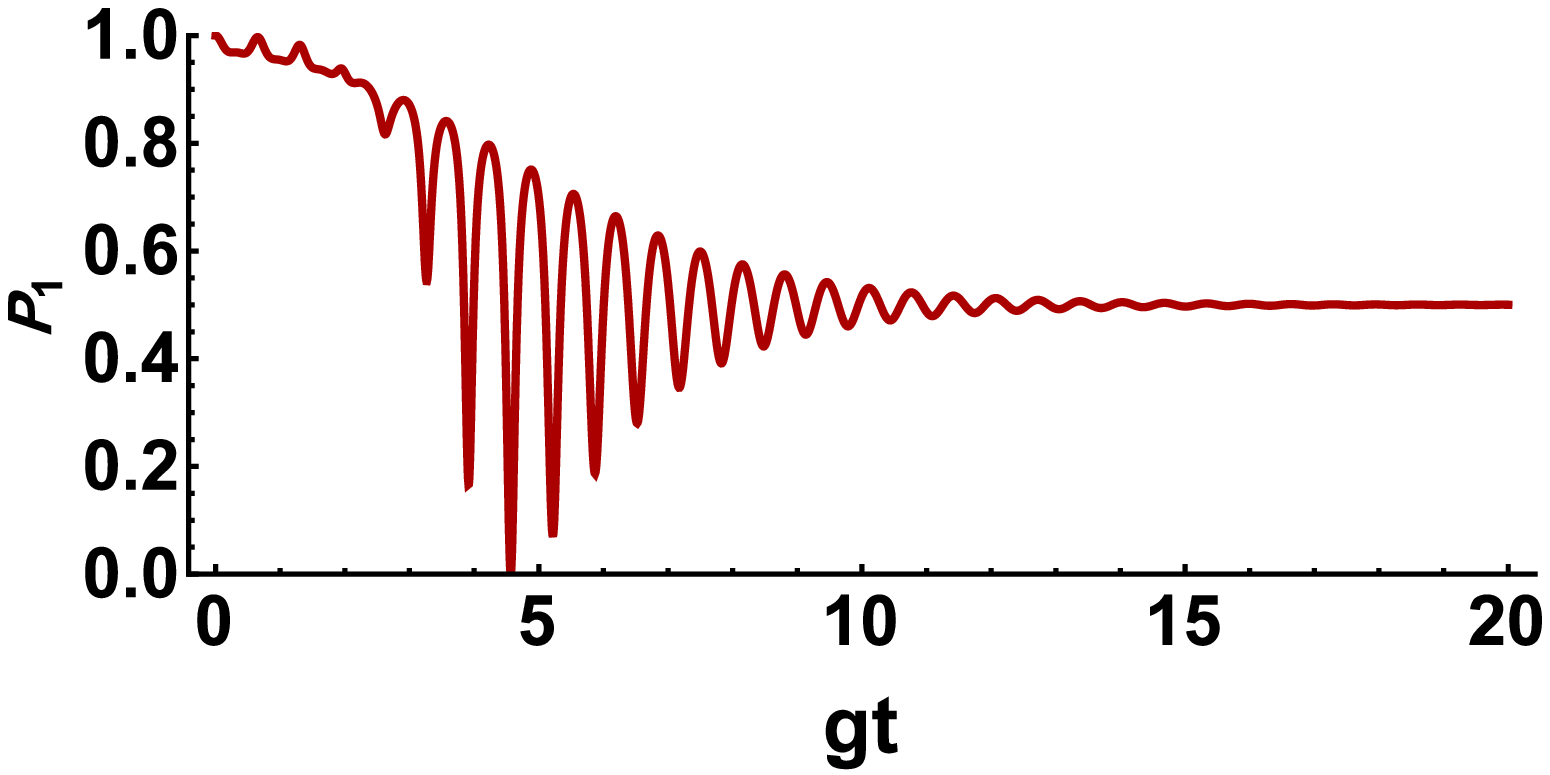}}
\hspace{0.05\textwidth}
\subfigure[\label{fig.1p} \ $\frac{\Delta}{g}=7$, $\frac{\chi}{g}=0$, $\frac{\kappa}{g}=0.1$, $\frac{\varGamma}{g}=0.3$]{\includegraphics[width=0.35\textwidth]{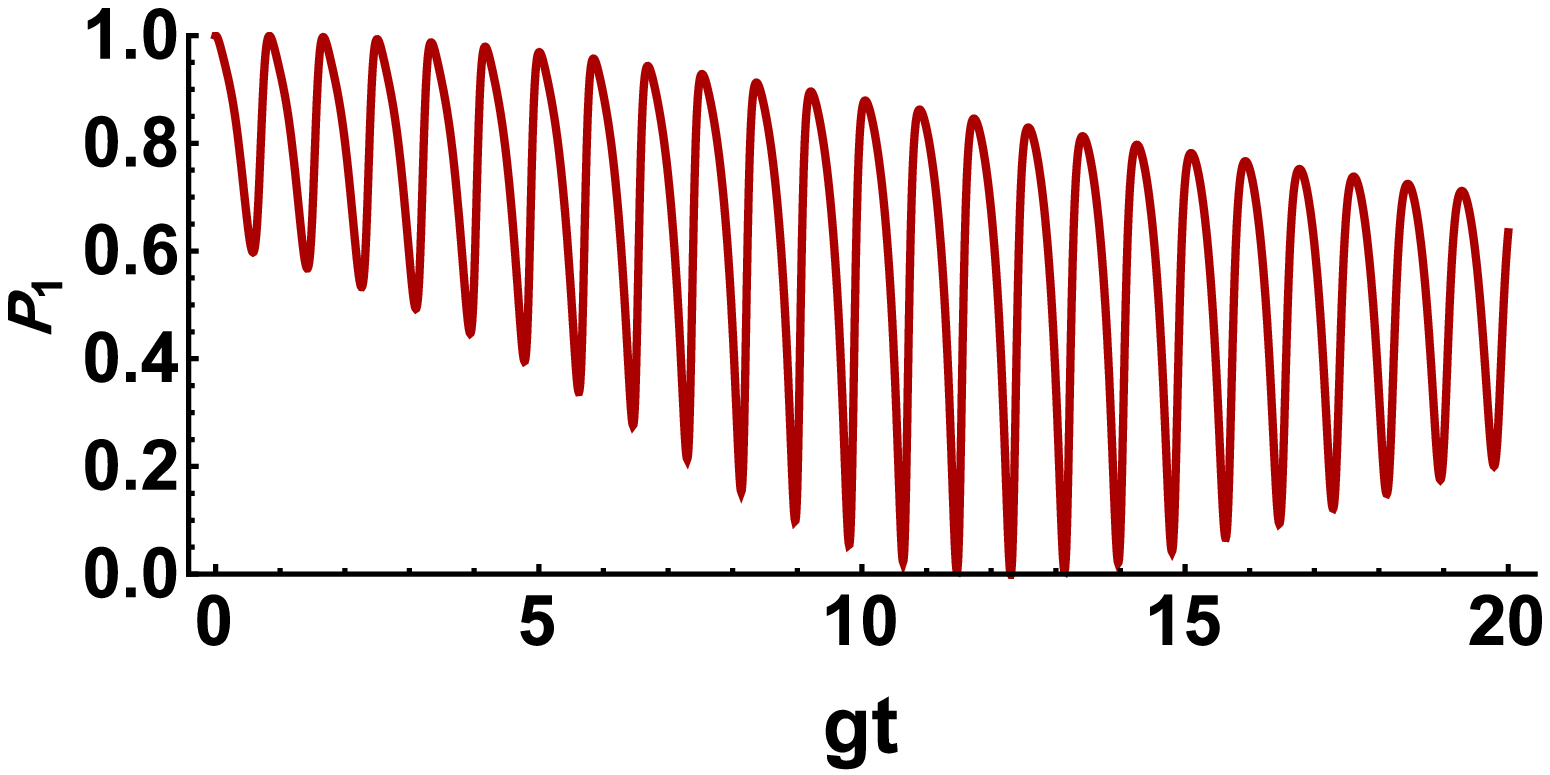}}
\hspace{0.05\textwidth}
\subfigure[\label{fig.22} \ $\frac{\Delta}{g}=7$, $\frac{\chi}{g}=0.7$, $\frac{\kappa}{g}=2$, $\frac{\varGamma}{g}=3$]{\includegraphics[width=0.35\textwidth]{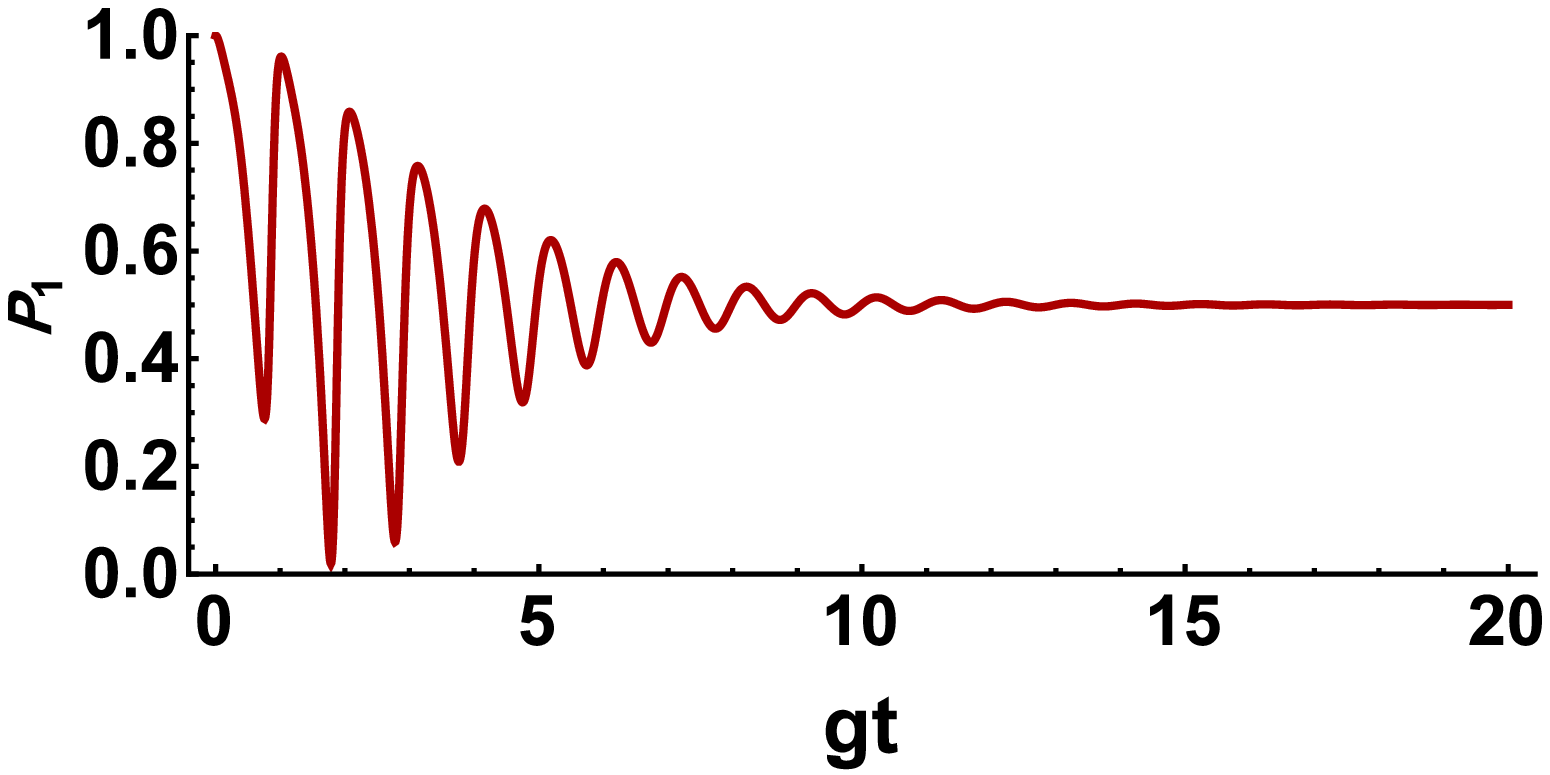}}
\caption{\label{fig.ocprob} The time evolution of occupation probability after entanglement swapping as function of the scaled time $gt$, for (a), (b) $\theta=\frac{\pi}{4}$, $\phi=0$ (c), (d) $\theta=\frac{\pi}{3}$, $\phi=\frac{3\pi}{2}$.}
\end{figure}

Let us now consider the case in which  $\theta=\frac{\pi}{4}$, $\phi=0$ and study the entanglement of atom-atom state (Eq. (\ref{densityoperator})). Figure \ref{fig.tpi4} illustrates the concurrence as a function of the scaled time $gt$ for different parameters. In Fig. \ref{fig.n5} the detuning and Kerr parameters are zero ($\frac{\Delta}{g}=\frac{\chi}{g}=0$) and the atomic decay and photon scape rates are equal, ($\frac{\kappa}{g}=\frac{\varGamma}{g}$) where we mentioned that under this condition our system converts to ideal system. Fig. \ref{fig.n5} has regular periodic behaviour and the concurrence reaches to 1 in some instants of time according to Eq. (\ref{Tnteta0}) and in these times atom-atom state behaves as the Bell state. In Fig. \ref{fig.nn9}, by increasing $\Delta/g$ to 10, the maxima of entanglement have been decreased, in fact, the detuning parameter has badly effect on the entanglement. In Fig. \ref{fig.10} in comparison with Fig. \ref{fig.nn9}, $\chi/g$ has been increased and we can see that the maxima of entanglement has not been achieved. But in Fig. \ref{fig.11}, in resonance condition ($\Delta=0$), by increasing $\chi/g$, entanglement has been reached to 1 and again atom-atom state converts to the Bell state. In comparison with Fig. \ref{fig.n5}, by increasing the effect of Kerr medium, the time period of entanglement has been increased. In Fig. \ref{fig.nn} we increase the amount of $\chi/g$ to 1, the behaviour of Fig. \ref{fig.nn} is regularly periodic and at times denoted by Eq. (\ref{Tnteta0}) the concurrence has reached to 1 and in comparison with Figs. \ref{fig.n5} and \ref{fig.11} entanglement in more times may be observed. These figures (\ref{fig.n5}-\ref{fig.nn}) have been plotted when $\kappa=\varGamma$. In Fig. \ref{fig.n6} dissipation has been added ($\kappa\neq\varGamma\neq0$) where a chaotic behaviour is seen. In comparison with Fig. \ref{fig.nn9} (where $\kappa=\varGamma$), we can find that the maxima of entanglement in greater times has been occurred. In Fig. \ref{fig.13} in addition to the presence of dissipation, the Kerr medium has been added. In this figure in comparison with Fig. \ref{fig.10}, we can see that the maxima of entanglement have been achieved, the regular periodic behaviour of concurrence has been destroyed and the concurrence has been fixed to 1 at more intervals of times.\\
 In Fig. \ref{fig.conp} we have changed the initial atomic state to $\theta=\frac{\pi}{3}$, $\phi=\frac{3\pi}{2}$ and considered the concurrence measure (Eq. (\ref{cont})). In Fig. \ref{fig.1} the detuning and Kerr parameters are taken to be zero ($\frac{\Delta}{g}=\frac{\chi}{g}=0$) and $\kappa=\varGamma$, hence our system is as an ideal system. The temporal behaviour of Fig. \ref{fig.1} is regularly periodic and the concurrence in some times (according to Eq. (\ref{Tnteta0})) being equal to 1. In fact Fig. \ref{fig.1} has two maxima (a relative minimum 0.6 and an absolute maximum 1). In Fig. \ref{fig.5} in the absence of Kerr medium and in the presence of detuning, by increasing $\Delta/g$, the maxima of entanglement have been decreased, but in comparison with Fig. \ref{fig.1} the concurrence has reached to an average of the two maxima of Fig. \ref{fig.1} (i.e. 0.8) in more instants of time. In Fig. \ref{fig.nn5} by decreasing $\Delta/g$, in comparison with Fig. \ref{fig.5}, and by increasing $\chi/g$, entanglement has been reached to 1 and in some times (according to Eq. (\ref{Tnteta0})) atom-atom state has transformed to  Bell state. In Fig. \ref{fig.6} in comparison with Fig. \ref{fig.1} the Kerr medium effect has been increased and we can see from this figure that entanglement has been reached to its maxima in more times. In Fig. \ref{fig.ll} in addition to the presence of detuning, dissipation parameters are not equal (in fact for the case in which $\kappa\neq\varGamma\neq0$) and in comparison with Fig. \ref{fig.5} the concurrence has been reached to 1 but the regular periodic behaviour of concurrence has been destroyed. In Fig. \ref{fig.2} we consider the effect of dissipation and Kerr parameters. One can see that the chaotic behaviour of the concurrence tends to 1 as time goes on.\\
In Fig. \ref{fig.ocprob} we consider the condition under which the concurrence reaches to 1 (when $P_1(t)=P_2(t)=0.5$ in Eq. (\ref{eqp})). Figs. \ref{fig.p1} and \ref{fig.1313} have been plotted for the conditions of Figs. \ref{fig.n6} and \ref{fig.13}, respectively. We can see from these figures that when $P_1(t)$ is being equal to $0.5$ the concurrence reaches to 1. Figs. \ref{fig.1p} and \ref{fig.22} in comparison with Figs. \ref{fig.ll} and \ref{fig.2}, respectively illustrates that at the times which occupation probability has reached to $0.5$, maximally entanglement of the two qubits has been achieved.

\newpage
 \section{Summary and conclusions}\label{sec.con}

In this work we have considered two independent nonperfect cavities each consists of a decaying atom interacting with a single-mode dissipative field  in the presence of a Kerr medium.  The description of the model is governed by JCM Hamiltonian which produces atom-field entangled state in each cavity. We have implemented the entanglement swapping protocol to transform the entanglement from the two atom-field subsystems to the qubit-qubit system by an interference measurement performed on the fields leaving the cavities (BSM).
We have discussed the influences of initial atom-field state, detuning parameter, Kerr medium and dissipation effects on the atom-atom swapped entanglement.
We have found that the maximum entanglement can be created between the two atoms with performing appropriate circumstances. In particular, we have demonstrated that whenever $P_1(t)=P_2(t)=0.5$ holds, the state of atom-atom is projected into a maximally entangled state. Accordingly, we have obtained the two following conditions: first is $\theta=(2n+1)\pi/2$, that the maximum of entanglement is independent of time (for any values of $\chi$, $\Delta$, $\kappa$ and $\varGamma$), second condition is $ \left|  \eta(t)\right|  ^4-\frac{64g^4}{\Omega_0^4}\sinh^4\left( \Omega_0 t/2\right)=0$ for $\kappa=\varGamma$  which is independent of $\theta$. In the first case, the state of the atom-atom is always projected into the Bell state $\ket{\Psi^-}$. While, going back with the second condition, the state of atom-atom can be projected into $\ket{\Psi^-}$ or $\ket{\Psi^+}$. In this case, we exactly determined the state of atom-atom as a function of $\theta$ (see Fig. \ref{fig.Theta}).\\
We also found that if the decay rate and photon scape rate be equal, in fact, our system behaves as an ideal system and in this condition the entanglement (or concurrence) possesses regular periodic. But in the presence of dissipation ($\kappa\neq\varGamma\neq0$) the behaviour of concurrence and occupation probability is not periodic and in more instants of time the two-qubit state is converted to maximally entangled state. Also, by increasing the detuning parameter in the absence of dissipation (or equivalently when $\kappa=\varGamma$, see Eq. (\ref{landa})) and Kerr medium the maximum of concurrence has been decreased.


\end{document}